\documentclass[prb,showpacs,twocolumn,floatfix]{revtex4}
\usepackage{graphicx,bm}

\begin{document}

\title{Constructing smooth potentials of mean force, radial
  distribution functions and probability densities from sampled data}

\author{Ramses van Zon and Jeremy Schofield}

\affiliation{Chemical Physics Theory Group, Department of Chemistry,
  University of Toronto, 80 Saint George Street, Toronto, Ontario M5S
  3H6, Canada}

\date{March 3, 2010}

\begin{abstract}
In this paper a method of obtaining smooth analytical estimates of
probability densities, radial distribution functions and potentials of
mean force from sampled data in a statistically controlled fashion is
presented.  The approach is general and can be applied to any density
of a single random variable.  The method outlined here avoids the use
of histograms, which require the specification of a physical parameter
(bin size) and tend to give noisy results.  The technique is an
extension of the Berg-Harris method [B.A.\ Berg and R.C. Harris,
  Comp.\ Phys.\ Comm.\ \textbf{179}, 443 (2008)], which is typically
inaccurate for radial distribution functions and potentials of mean
force due to a non-uniform Jacobian factor.  In addition, the standard
method often requires a large number of Fourier modes to represent
radial distribution functions, which tends to lead to oscillatory
fits.  It is shown that the issues of poor sampling due to a Jacobian
factor can be resolved using a biased resampling scheme, while the
requirement of a large number of Fourier modes is mitigated through an
automated piecewise construction approach.  The method is demonstrated
by analyzing the radial distribution functions in an
energy-discretized water model.  In addition, the fitting procedure is
illustrated on three more applications for which the original
Berg-Harris method is not suitable, namely, a random variable with a
discontinuous probability density, a density with long tails, and the
distribution of the first arrival times of a diffusing particle to a
sphere, which has both long tails and short-time structure.  In all
cases, the resampled, piecewise analytical fit outperforms the
histogram and the original Berg-Harris method.
\end{abstract}


\maketitle

\section{Introduction}

In many situations the probability to find a particle at a certain
distance from a given other particle is of interest.  This probability
is given by the radial distribution function.  In addition to
providing detailed information on the local structure of a system, one
can often express thermodynamic quantities such as the pressure,
energy and compressibility in terms of the radial distribution
functions.\cite{McQuarrie} Furthermore, the radial distribution
function can be reformulated in terms of a potential of mean force,
which is of great importance in multi-scale simulations that take the
potential of mean force as input in Langevin or Brownian dynamics
simulations.\cite{FrenkelSmit,Berendsen07}

Radial distribution functions are usually constructed in simulations
by forming a histogram of sampled inter-particle distances.
\cite{FrenkelSmit} However, histograms contain the bin size as a free
parameter, and can be rather noisy for a poorly selected value of this
parameter.  For probability densities, an alternative method recently
proposed by Berg and Harris avoids histograms and yields a smooth
analytical form for the probability density which describes the sample
data statistically at least as well as the noisy
histogram.\cite{BergHarris08} This method has already proved useful in
the context of the computation of quantum free energy differences from
non-equilibrium work distributions,\cite{VanZonetal08b} the
determination of the density in Bose-Einstein condensates,
\cite{Gerickeetal08} and in the distribution of a reaction coordinate
in simulations of chemical reactions.\cite{Monesetal09}

Having a similar method for potentials of mean force and radial
distribution functions would have many advantages.  As in the
Berg-Harris method, such an approach would avoid the noise that
accompanies the standard histogram method, while no \textit{a priori}
choice of bin size is required. Furthermore, a smooth radial
distribution would give a better representation of the potential of
mean force (which is related to the logarithm of the radial
distribution function), and allows the function to be evaluated at any
point in the range over which it is defined.  Finally, the expressions
for the pressure, energy and compressibility in terms of the radial
distribution functions involve integrals of the radial distribution
function over $r$.  Such integrals can be evaluated more accurately
from an analytic form than from a histogram, whose accuracy is
restricted by the bin size.

The purpose of this paper is to develop an approach to obtain smooth
radial distribution functions and potentials of mean force from
sampled inter-particle distances.  The resulting method turns out to
be suitable not just for radial distribution functions and potentials
of mean force, but for a large class of densities of single random
variables.

The paper is structured as follows.  Sec.~\ref{sec:testcase} presents
a model of water in which rigid molecules interact through a
discretized potential.  Construction of radial distribution functions
from data derived from this model will be used as a running test case.
This section also contains some details on the generation of sampled
data through simulation.  The Berg-Harris method for smoothing
probability densities in a statistically controlled fashion, and the
connection between probability distribution functions and radial
distribution functions, are briefly reviewed in Sec.~\ref{sec:review}.
In Sec.~\ref{sec:failure}, simulation results are presented that show
the shortcoming of the Berg-Harris method for radial distribution
functions. The extended method is developed in
Sec.~\ref{sec:extending}, with Sec.~\ref{sec:obstacles} containing an
explanation for the poor performance and its solution through
statistical resampling.  Sec.~\ref{sec:physicalproblem} addresses an
additional problem related to the particular shape of radial
distribution functions, which is solved by extending the method to use
piecewise analytic functions.  In Sec.~\ref{sec:furtherapp}, the
generality of the method is illustrated by presenting the results of
applying the method to data drawn from three different probability
densities which are problematic for the original Berg-Harris method.
The paper concludes with a discussion in Sec.~\ref{sec:conclusions}.

\section{System: a discrete water model}
\label{sec:testcase}

In the development of the smooth approximation method below, a model
of rigid water molecules subject to a discretized interaction
potential between the molecules will be used as a running test case
for the construction of radial distribution functions.  Since a water
molecule consists of two kinds of atoms (oxygen and hydrogen), there
are three radial distribution functions in this system,
$g_\mathrm{OO}$, $g_\mathrm{OH}$ and $g_\mathrm{HH}$, which turn out
to have quite different character and therefore give a more stringent
test of the smooth fitting methods than a single-atom model would.

The relative distances between the atomic sites in molecules are
fixed, making the molecules rigid bodies.  Each state of each body $i$
can therefore be described by its center-of-mass position $\mathbf
r_i$, its orientation or attitude matrix $\mathsf A_i$ that transforms
coordinates from the lab frame to a body-fixed frame for molecule $i$,
and the associated linear and angular momenta.  Here the body-fixed
frame is chosen so that the third row of the matrix $\mathsf A_i$
corresponds to the direction of the molecule's dipole $\bm\mu_i$,
whose magnitude is fixed at a value~$\mu$.

The interaction potential between a pair of molecules $i$ and $j$ is a
discrete version of the soft sticky dipole potential
\cite{LiuIchiye95,ChandraIchiye99,Tanetal03,FennellGezelter04}
\begin{equation}
  v_{ij} = v^\mathrm{LJ}(\mathbf r_i-\mathbf r_j)
+v^\mathrm{dp}(\mathbf r_i-\mathbf r_j, \mathsf A_i,\mathsf A_j)
+v^\mathrm{sp}(\mathbf r_i-\mathbf r_j, \mathsf A_i,\mathsf A_j),
\end{equation}
where the Lennard-Jones, dipole, and sticky parts of the potential
are, respectively, given by
\begin{eqnarray}
v^\mathrm{lj}(\mathbf r) &=&
4\epsilon\left[\left(\frac{\sigma}{r}\right)^{12}
-\left(\frac{\sigma}{r}\right)^{6}\right]
\\
v^\mathrm{dp}(\mathbf r, \mathsf A_i,\mathsf A_j) &=& 
\frac{\bm\mu_i\cdot\bm\mu_j}{r^3}-\frac{\bm\mu_i\cdot\mathbf r\ 
\bm\mu_j\cdot\mathbf r}{r^5}
\\
v^\mathrm{sp}(\mathbf r, \mathsf A_i,\mathsf A_j) &=& 
\frac{v_0}{2}\Bigg[
s(r)\{\sin\theta_{ij}\sin2\theta_{ij}\cos2\phi_{ij}
\nonumber\\&&\qquad\quad
+\sin\theta_{ji}\sin2\theta_{ji}\cos2\phi_{ji}\}
\nonumber\\
&&\quad+s'(r)\{
 (\cos\theta_{ij}-0.6)^2(\cos\theta_{ij}+0.8)^2
\nonumber\\&&\qquad\qquad
+(\cos\theta_{ji}-0.6)^2(\cos\theta_{ji}+0.8)^2
\nonumber\\&&\qquad\qquad
-2 \omega^0
\}
\Bigg]
\end{eqnarray}
where $\theta_{ij}$ and $\phi_{ij}$ are the conventional spherical
angles of the vector $\mathsf A_i\mathbf r$, and $\theta_{ji}$ and
$\phi_{ji}$ are those of $\mathsf A_j\mathbf r$. Finally, the
switching function $s(r)$ is defined as
\begin{equation}
  s(r) = \left\{
  \begin{array}{ll}
    1 &\mbox{ for } r<r_L\\
    \frac{(r_U-r)^2(r_U+2r-3r_L)}{(r_U-r_L)^3} &\mbox{ for } r_L<r<r_U\\
    0 &\mbox{ for } r>r_U
  \end{array}
\right. ,
\end{equation}
while $s'(r)$ is given by the same form with primed parameters $r'_L$
and $r'_U$.

Here, the so-called SSD/E reparameterization of the model was used,
for which the parameter values are $\epsilon=0.152$ kcal/mol,
$\sigma=3.035$ \AA, $\mu=2.42$ D, $v_0=3.90$ kcal/mol, and
$\omega^0=0.07715$, while the cut-off parameters in the functions $s$
and $s'$ are taken to be $r_L=2.4$ \AA, $r_U=3.8$ \AA, $r_L'=2.75$
\AA\ and $r_U'=3.35$ \AA.\cite{FennellGezelter04}

The discontinuous interaction potential in our system is obtained from
the smooth potential by the controlled energy discretization method
presented in Ref.~\onlinecite{VanZonSchofield08}. In this
discretization, a cut-off naturally arises, and therefore no reaction
field was included.

The natural simulation technique for systems with discretized
potentials is discontinuous molecular dynamics, or DMD.
\cite{dmd1,dmd2,VanZonSchofield08} In DMD, the dynamics of the system
is free (no forces or torques are present) in between interaction
events at which linear and angular momenta change.  By its very
nature, the energy-discretization scheme used in DMD is symplectic,
time-reversible and strictly conserves the total (discretized) energy.
It has no fixed time-step, but moves from event to event. The event
frequency, which sets the efficiency of the method to a large degree,
is determined by the level of discretization of the potential energy:
the finer the discretization, the more events occur.  One typically
finds that a discretization of the order of $\frac12 k_BT$ already
suffices for a reasonably accurate simulation of the smooth
system.\cite{VanZonSchofield08} In the simulations of which the
results are presented below, the discretization of the potential
energy was set to $\frac14k_BT$.

The orientational dynamics of free rigid bodies is an important aspect
in the calculation of interaction times.\cite{dmd1,dmd2} The solution
of the equations of motion for the angular momenta and attitude matrix
$\mathsf A$ depends on the symmetry of the body through the principal
moments of inertia.  Although any value of the principal moments can
be utilized to sample configurations of a system of rigid molecules,
the DMD simulations presented here made use of the exact solution of
the equations of motion for a free, asymmetric rigid
rotor,\cite{dmd1,VanZonSchofield07a} and hence were based on the exact
dynamics of the system.

At the time of a distance measurement in the simulation, the forces
and torques are most likely zero, since these quantities are non-zero
only at discrete time points.  This makes alternative smoothing
methods, such as the weighted residual method\cite{CyrBond07} and
methods based on equilibrium identities using (smooth)
forces,\cite{AdibJarzynski05,BasnerJarzynski08} not applicable
here. Since only the inter-particle distances are available as input
for the determination of the analytically-fitted radial distribution
functions, the comparison with histogram-based radial distribution
functions is more equitable.

\section{Review}
\label{sec:review}

\subsection{Smoothing probability densities with the Berg-Harris method}
\label{sec:spd}

Consider a random variable $r$ which has a probability density $p(r)$.
Suppose that one has a sample of $n$ data points $\{r_i\}$ that are
independently drawn from the density $p(r)$.  How can one estimate
$p(r)$ from the data points? One way is to bin the data points into a
histogram.  Because histograms are quite sensitive to statistical
noise, Berg and Harris developed the following alternative procedure
to obtain an analytical estimate for the probability density from the
data.\cite{BergHarris08}

In the Berg-Harris method, the data are first sorted such that
$r_{i}<r_{i+1}$.  Using the sorted data, the empirical cumulative
distribution function $\bar F$ is defined in the range $[r_1,r_n]$ as
\begin{equation}
\bar F(r) = \frac{i}{n} \qquad \mbox{for $r_i \leq r < r_{i+1}$}.
\label{ecdf}
\end{equation}
Although $\bar F$ becomes a better approximation to the true
cumulative distribution $F(r)=\int^r p(r')\,dr'$ with increasing
sample size $n$, it is a function with many steps for any finite value
of $n$ so that its derivative is not analytic but rather consists of
delta functions.

The next step consists of writing the function $\bar F$ as a sum of a
linear term and a Fourier expansion. The expansion is truncated at the
$m$th term:
\begin{equation}
\bar F(r) \approx F_m(r) \equiv F_0(r) + \sum_{j=1}^m d_j \sin[j\pi F_0(r)],
\label{approx}
\end{equation}
where the linear term is defined as
\begin{eqnarray}
F_0(r) &=&\frac{r-r_1}{r_n-r_1}.
\label{xdef}
\end{eqnarray}
Furthermore, the Fourier coefficients $d_j$ in Eq.~(\ref{approx}) are
determined from
\begin{equation}
d_j = \frac{2}{r_n-r_1} \int_{r_1}^{r_n} [\bar F(r)-F_0(r)] 
 \sin[j\pi F_0(r)]\:dr.
\end{equation}
When $\bar F$ is approximated by $F_m$, the probability density $p$ is
approximately
\begin{equation}
p_m(r) = 
\frac{1}{r_n-r_1} \left\{1 
+ \pi \sum_{j=1}^m d_j j\cos[j\pi F_0(r)]\right\}.
\label{papprox}
\end{equation}
Because $\bar F-F_0$ is zero at the end points of the interval
$[r_1,r_n]$, and the Fourier modes form a complete orthonormal basis
of the space of such functions, the empirical cumulative distribution
$\bar F$ and the associated probability density are reconstructed
exactly by Eqs.~(\ref{approx}) and (\ref{papprox}) as
$m\to\infty$. However, as mentioned above, this limit would yield a
series of delta functions for the probability density
$\bar{p}=p_\infty$.  The aim is to truncate the series at a level $m$
which is not too high that one is fitting the noise, but high enough
to give a good smooth approximation to $\bar F$ and $\bar{p}$.

The final step of the procedure is therefore to find the appropriate
number of Fourier terms $m$ in a statistically controlled fashion. The
value of $m$ is determined here using the Kolmogorov-Smirnov (K-S)
test.\cite{BergHarris08,NumRecipes} This test determines how likely it
is that the difference between the empirical cumulative distribution
function $\bar F$ and its analytical approximation $F_m$ is due to
noise.\cite{footnote1} The test takes the maximum variation $D_m$
between $\bar F$ and $F_m$ over the sampled points
($D_m=\max_{i=1\dots n}|F_m(r_i)-\bar F(r_i)|$), and returns a
probability $Q_m=Q(D_m)$ that the difference between the two
cumulative distribution functions is due to chance. The functional
dependence of $Q$ on $D$ is known to a good (asymptotic)
approximation.\cite{Stephens70}

A small value of $Q_m$ indicates that the difference between the
cumulative distribution functions is statistically significant,
i.e.\ the quality of the expansion $F_m$ is insufficient to represent
the data.  One therefore carries out a process of progressively
increasing the number of Fourier modes $m$ and evaluating $F_m$ as
well as $Q_m$ until the value of $Q_m$ is larger than some convergence
threshold $Q_\mathrm{cut}$. A reasonable value for this convergence
value is $Q_\mathrm{cut}=0.6$.

In Ref.~\onlinecite{VanZonetal08b}, errors were estimated using the
jackknife algorithm,\cite{Efron79} and the same method will be used
here.  However, since the inter-particle distance data naturally come
in blocks, each corresponding to a single configuration, not all data
points are independent. To account for this, we use a block-version of
the jackknife method, in which a single block of data is omitted in
each jackknife sample.\cite{Kuensch89}

\begin{figure*}
\centerline{\includegraphics[width=.33\textwidth]{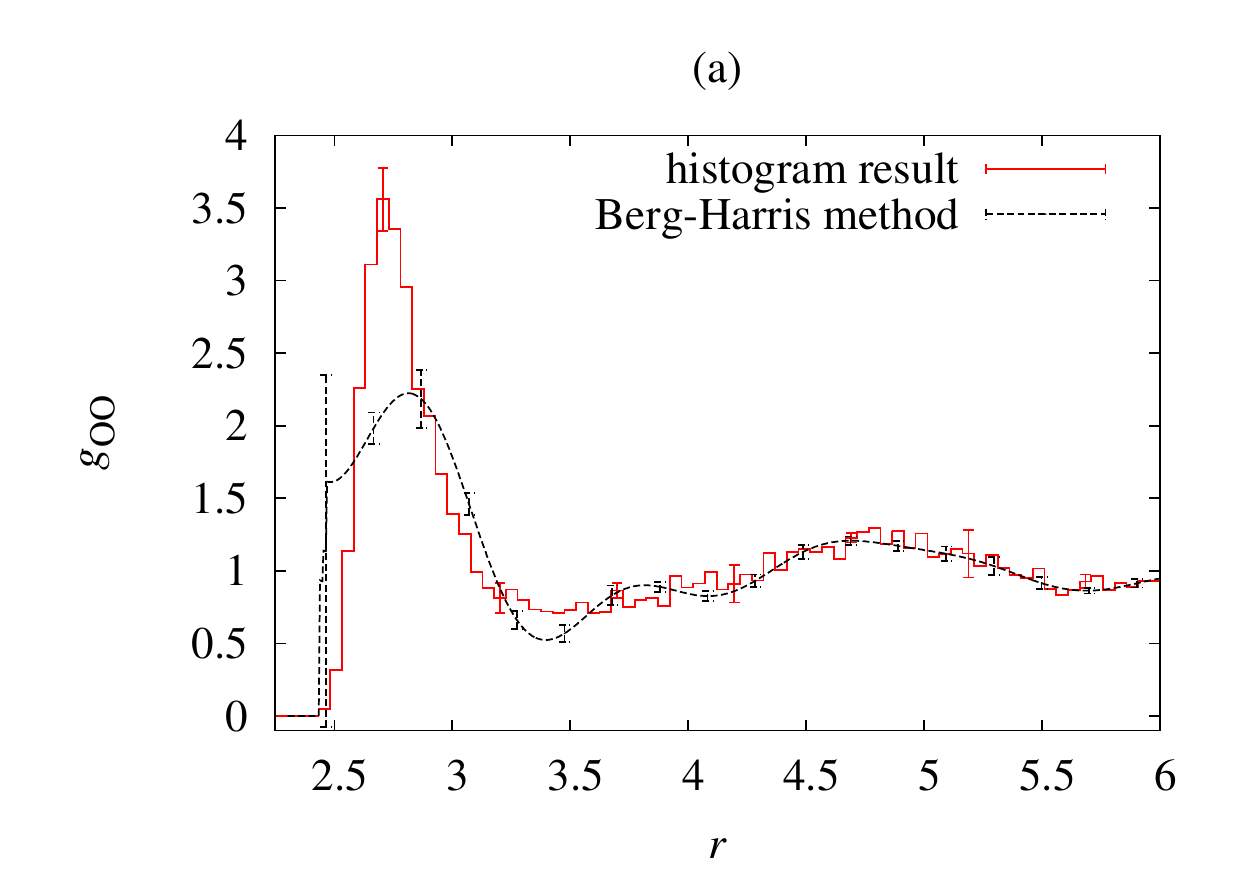}%
            \includegraphics[width=.33\textwidth]{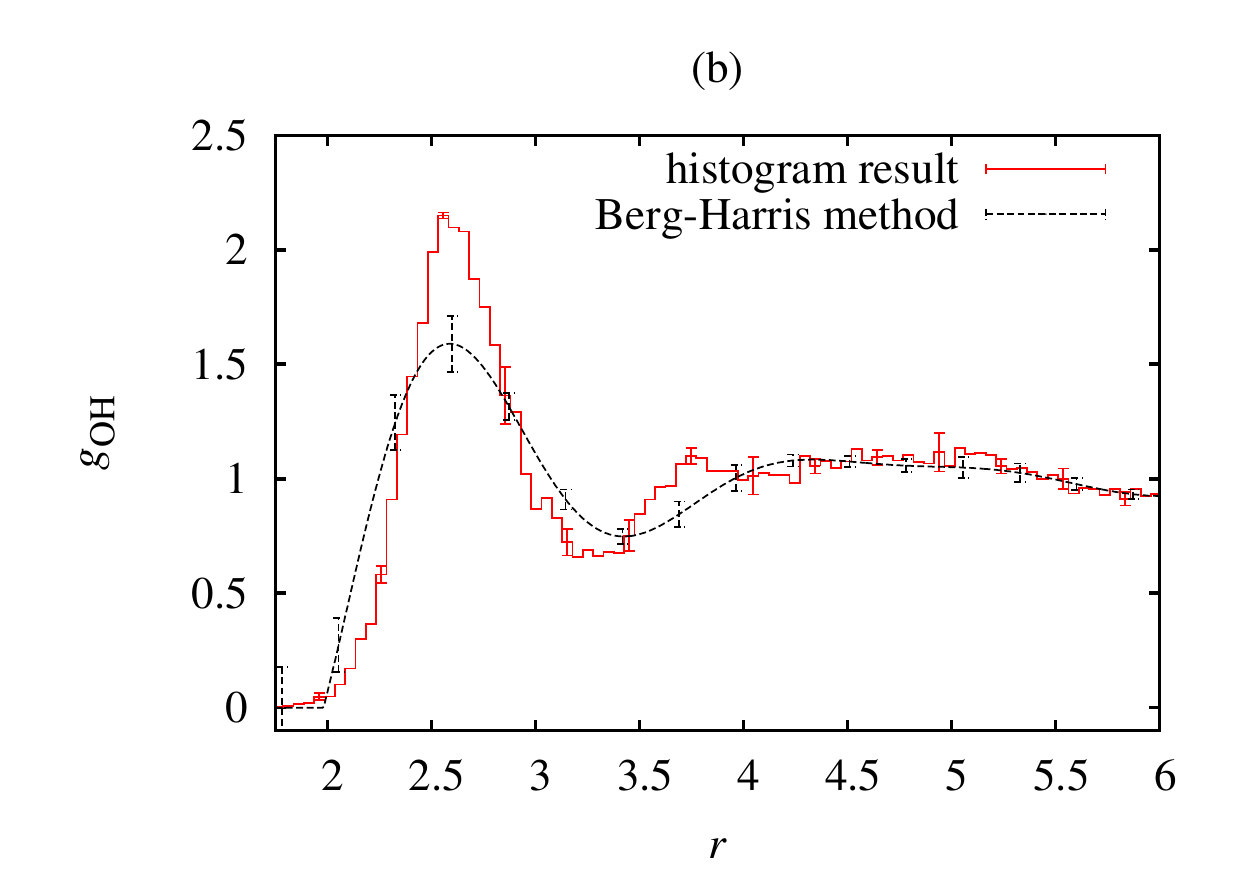}%
            \includegraphics[width=.33\textwidth]{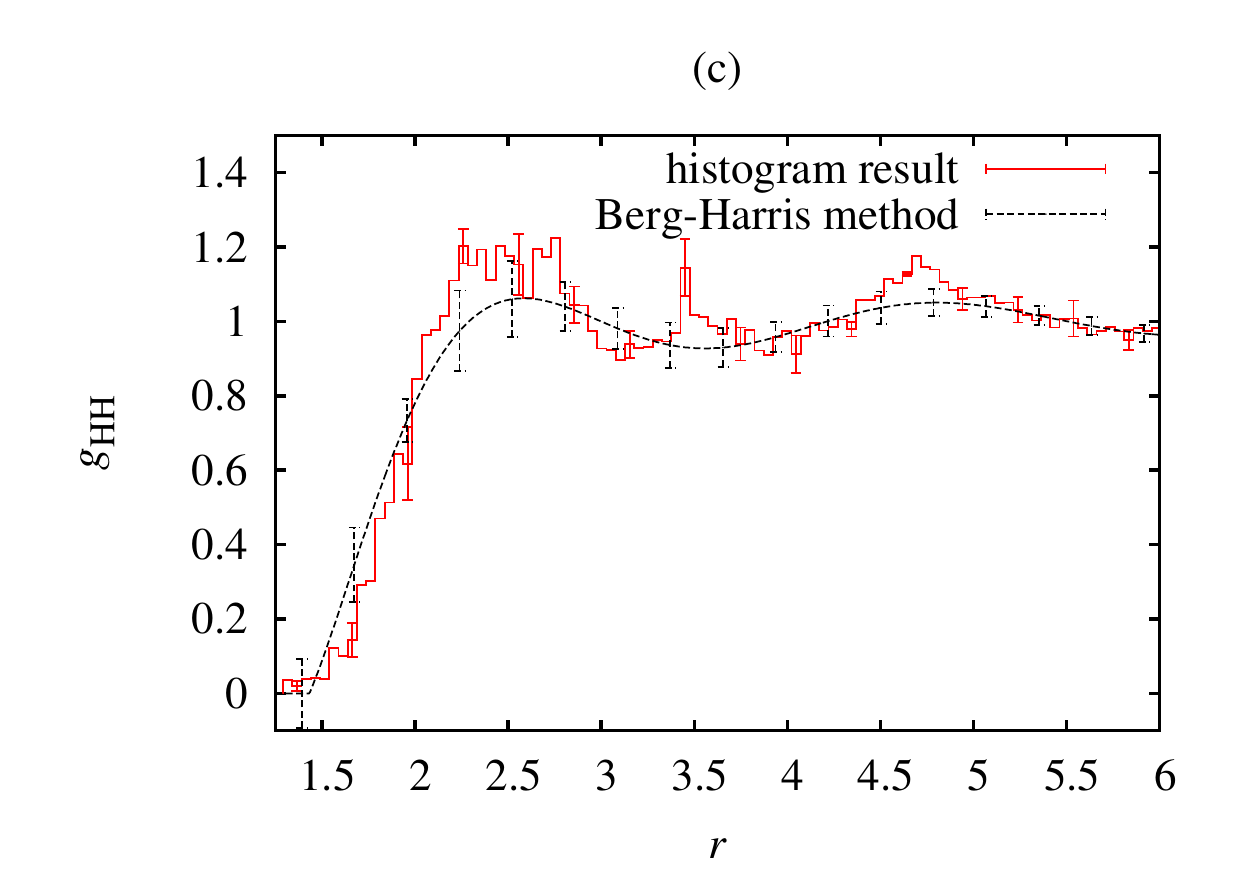}}
  \caption{Performance of the straightforward application of the
    Berg-Harris smoothing methods applied to the radial distribution
    functions $g_\mathrm{OO}$ (a), $g_\mathrm{OH}$ (b), and
    $g_\mathrm{HH}$ (c) of the discretized water model.  For
    comparison, the results from the histogram method are also shown
    (with error bars representing 95\% confidence intervals at
    selected points).  Here the range over which the radial
    distribution functions are analyzed is defined to be $[0,L/2]$,
    with $L/2 = 12.4$ \AA}
  \label{fig:1}
\end{figure*}

\subsection{From probability densities to potentials of mean force}
\label{sec:rdfpd}

To be able to apply the above smoothing method to radial
distributions, for which histograms are presently the method of
choice, one has to make the connection between a probability density
on the one hand, and the radial distribution function and potentials
of mean force on the other hand. This connection will be briefly
reviewed here because some of the details are needed below.

Consider a system of a single type of particle, in which the number of
particles is $N$ and the volume of the system is $V$.  The radial
distribution function, denoted by $g(r)$, is the density of particles
at a distance $r$ away from a chosen first particle relative to the
mean density $N/V$. The true density at a distance $r$ from a first
particle is thus $(N/V)g(r)$. The mean number of particles at a
distance between $r$ and $dr$ from a given particle is then
$n(r)=(N/V)g(r)4\pi r^2dr$. The Jacobian factor $4\pi r^2$, which is,
of course, the surface area of a sphere with radius $r$, will play an
important role below.  The probability of a particle being at a
distance $r$ is equal to the number of particles with this $r$,
divided by the total number of particles, $N$,
i.e. $p(r)dr=n(r)/N=g(r)4\pi r^2dr/V$, so one has the relation
\begin{equation}
g(r) = \frac{V}{4\pi r^2}\, p(r).
\label{gp}
\end{equation}

Systems with different types of particles give rise to different
radial distribution functions $g_{ij}(r)$, where $i$ and $j$ label the
kinds of particles.  A similar argument to that above leads to the
relation,
\begin{equation}
g_{ij}(r) = \frac{V}{4\pi r^2}\, p_{ij}(r),
\label{gij}
\end{equation}
where $p_{ij}(r)$ is the probability density of distances between a
particle $j$ and a particle $i$.  Once $g_{ij}(r)$ is known, the
potential of mean force $\Phi_{ij}$ is found from\cite{McQuarrie}
\begin{equation}
\Phi_{ij}(r) = -k_BT\ln g_{ij}(r),
\label{pmf}
\end{equation}
where $T$ is the temperature of the system and $k_B$ is Boltzmann's
constant.

A further consideration for the applicability of the Berg-Harris
method to radial distribution functions is whether the K-S test may be
used at all. The K-S test assumes that the samples are independently
drawn.  Correlation between the samples may result in a bias in the
radial distribution functions.  Since nearby particles in an
instantaneous configuration of the system are correlated, this is
potentially an issue. If, however, the system is sufficiently large,
only a small fraction of the samples in a single configuration will be
spatially correlated, making the K-S test applicable to a very good
approximation. Furthermore, if one ensures that the configurations are
taken from the simulation at sufficiently large time intervals, time
correlations do not pose a problem either.

\section{Poor performance of the straightforward smoothing applied to
  radial distribution functions}
\label{sec:failure}

Using the Berg-Harris method to get a smooth probability densities
$p(r)$ from a sample of inter-particle distances, one expects to
obtain a good, smooth fit to $g(r)$ by using Eq.~(\ref{gij}).  To test
this expectation, the system of pure water in which rigid water
molecules interact via a discretized potential energy derived from the
soft sticky dipole model described in Sec.~\ref{sec:testcase} was
simulated.  For all simulation results presented in this section, the
parameters of the water model were as follows. The temperature is set
at $T=298~\rm K$, the number of particles is $N=512$, the cubic
simulation box has sides of length $L=24.8$\AA, so that the density is
$1.0$ kg$/$l.  The principal moments of inertia of a rigid water
molecule are $I_x = 0.0337365 \, m_{H_2O}$\AA$^2$, $I_y = 0.0635040 \,
m_\mathrm{H_2O}$\AA$^2$, and $I_z = 0.0972405 \,
m_\mathrm{H_2O}$\AA$^2$, where $m_\mathrm{H_2O}$ is the molecular mass
of water. After equilibration, the simulations were run for $8$
picoseconds, in which the inter-particle distances were sampled every
$2$ picoseconds (long enough for the system to decorrelate), for a
total of $4$ configurations.

From these data, the radial distributions $g_\mathrm{OO}$,
$g_\mathrm{OH}$ and $g_\mathrm{HH}$ were determined in the simulations
through histograms and by following the smoothing procedure of
Sec.~\ref{sec:spd}, using a value for $Q_\mathrm{cut}$ of $0.6$. The
results for the three radial distribution functions are shown in
Fig.~\ref{fig:1}. Clearly, the two methods do not agree very well at
short inter-particle separations, despite the fact that the smooth
distance distributions are statistically good descriptions of the
distance data according to the K-S test.  In particular, note that the
peak heights defining the first solvation shell are not
well-described.

\section{Extending the Berg-Harris method}
\label{sec:extending}
Given these unsatisfactory results, the histogram method would be
preferred over the straightforward application of the Berg-Harris
method. But there is a way to fix the smoothing method to the extent
that the smoothing method becomes preferable over the histogram
method.  To understand how to fix the problem, one first needs to
understand its underlying causes.

\subsection{Over-represented large distances}
\label{sec:obstacles}

\subsubsection{Problematic Jacobian}

The convergence criterion of the smooth approximation relies on the
K-S test.\cite{BergHarris08,NumRecipes} This test is based only on the
maximum variation $D_m$ between $\bar{F}$ and $F_m$ and is more
sensitive to typical data points than to outliers.
Although this may appear to be a contradiction at first
glance, it is important that the K-S test depends on the maximum
deviation in the cumulative distribution function, rather than on the
deviation of the random variable from the mean.  To see that outliers
do not constitute very deviant points in the cumulative distribution,
suppose that, by chance, an outlier $x_{out}$ from the far left tail
of a distribution is found in a sample of size $n$. This would yield
an increase of magnitude of $1/n$ for the empirical cumulative
distribution $\bar F$ at $x_{out}$, while the cumulative distribution
$F(x)$ is practically zero since $x$ is in the region where the
distribution is very small. If $F_m$ is not too bad an approximation
to $F$, $\bar F-F_m$ will also be of order $1/n$ at $x_{out}$.
However, the deviation between $F_m$ and $\bar F$ at other points $x$
depends more on the goodness of fit than on the sample size, at least
for points $x$ in regions that contain a sizable fraction of the
samples: these are the `typical points'. For these points, there is
little sample size dependence, so that to first order in $n$, one has
$\bar F-F_m= O(n^0)$.  Since $O(n^{-1}) < O(n^0)$ for large enough
sample sizes, the outlier $x_{out}$ will not be seen as the most
deviant point in the K-S test, but rather, some typical value $x$ will
have the most deviant cumulative distribution.\cite{footnote1a}

The above argument holds when the quality of the fit in the typical
region is poor.  However, as the number of Fourier modes used to fit
$\bar F$ is increased, the quality of the fit of the cumulative
distribution $F$ improves in the typical region, and the maximum
deviation shifts to less probable values.  Because the probability
density function is small in the tails, this maximum deviation of the
distribution is not that large and the convergence criterion (the K-S
test) is met, resulting in a good fit in the typical region with a
poor fit in the tails.

The Berg-Harris method of Sec.~\ref{sec:spd} therefore works well for
probability densities without ``long tails,'' since typical values of
the variable will be the values of interest.  However, for radial
distribution functions, the focus on typical values is the origin of
the difficulties getting the straightforward Berg-Harris method to
work for radial distribution functions (cf.\ Fig.~\ref{fig:1}).
Typical values of $r$ are of no interest in radial distribution
functions.  In a homogeneous system such as a fluid, the typical
distances between any two particles is on the order of half the system
size, $L/2$.  The length scales of interest in the radial distribution
functions to describe local structure are typically much smaller than
that.

To make this point clearer, consider a dilute gas of hard spheres with
diameter $\sigma$.  In the limit of infinite dilution, the radial
distribution function $g(r)$ is zero for $r<\sigma$ and unity
otherwise. The probability density $p(r)$ of inter-particle distances
is then [cf.\,Eq.~(\ref{gp})]
\begin{equation}
p_\mathrm{dilute}(r) = \left\{\begin{array}{ll}
0          &\mbox{ for\ \ } r<\sigma \\
4\pi r^2/V &\mbox{ for\ \ } r\geq\sigma.
\end{array}\right.
\label{dilute}
\end{equation}
Because of the Jacobian factor of $4\pi r^2$, the most likely values
of $r$ are of the order of the largest possible $r$, which is $L/2$.
This means that in a sample of inter-particle distances in the dilute
hard sphere system, the large distance samples
($r_i=\mathcal{O}(L/2)$) are much more abundant than the small
distance samples ($r_i=\mathcal{O}(\sigma)$).  Thus, when
approximating $p(r)$ by a smooth function using the K-S test, the
long-distance part will be fitted very well, but the short-distance
behavior ($r=\mathcal{O}(\sigma)$) will not, because the test used is
sensitive to the typical values of $r$, which are $\mathcal{O}(L/2)$.

The conclusion that large distance values overwhelm the smaller ones
in which one is interested is valid beyond the dilute hard sphere
case, since the Jacobian factor in Eq.~(\ref{dilute}) acts on long
length scales and is therefore also present in systems of higher
density with non-negligible interactions.

\begin{figure}[t]
  \includegraphics[width=\columnwidth]{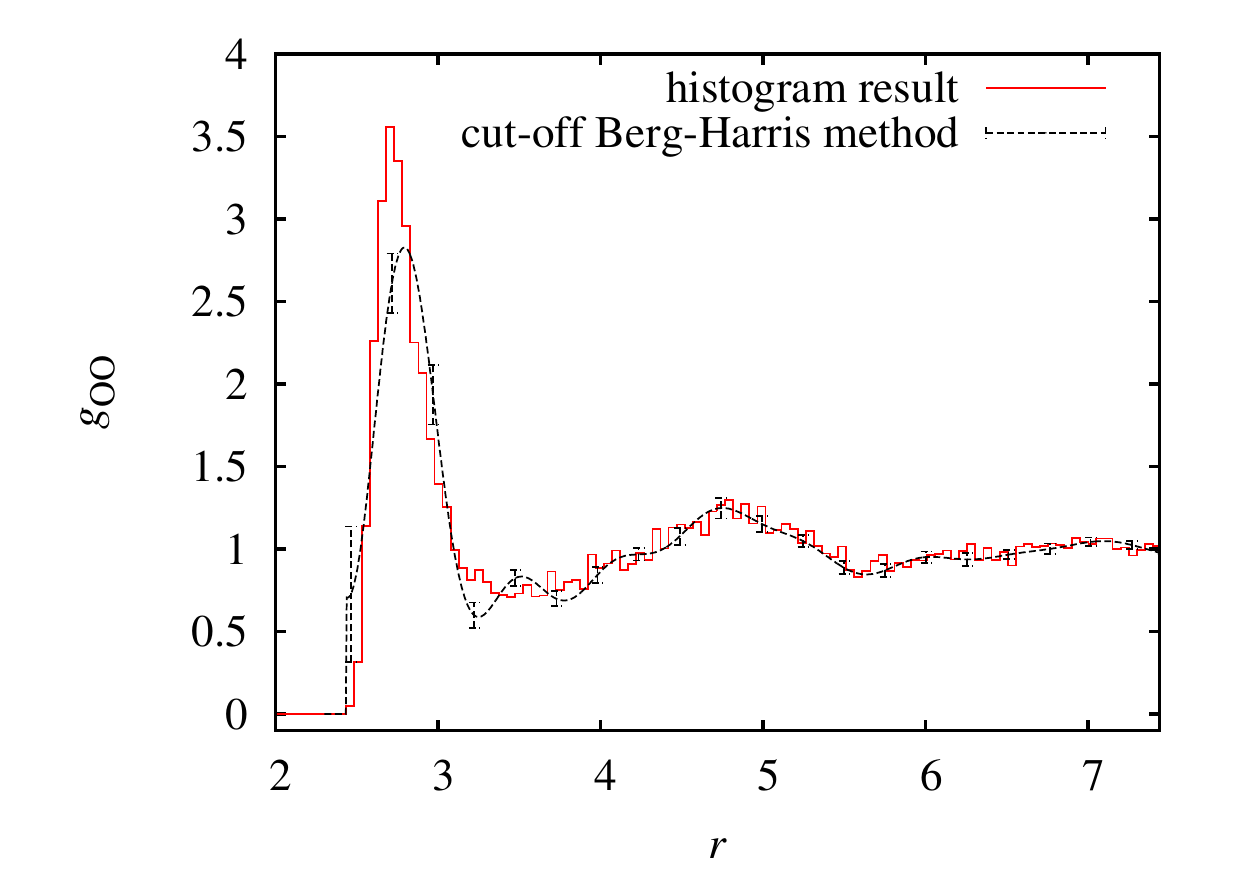}
  \caption{Result for $g_\mathrm{OO}$ of using a cut-off in the
    straightforward Berg-Harris smoothing, i.e., with the distance
    data restricted to $r<r_\mathrm{cut}=7.44$ \AA.  For comparison,
    the results from the histogram method are also shown (with error
    bars representing 95\% confidence intervals shown at
    selected~points).}
  \label{fig:2}
\end{figure}

A crude attempt at solving the large distance problem is to impose a
cut-off $r_{\rm cut}$ on the allowed values of $r$ in the sample of
inter-particle distances.  This cut-off should be of the order of the
distance over which $g(r)$ differs from one. But to determine that
distance one has to measure $g(r)$. Thus, one would have to guess a
value of the cut-off distance and adjust it until $g(r)$ approaches
one within a certain accuracy for $r<r_\mathrm{cut}$.  Unfortunately,
even when the cut-off is chosen in this way, the results, although
better than those in Fig.~\ref{fig:1}, are still not very impressive,
as Fig.~\ref{fig:2} for the oxygen-oxygen radial distribution function
shows with $r_\mathrm{cut}$ set to $7.44$ \AA (using again
$Q_\mathrm{cut}=0.6$).  One sees a lot of spurious oscillations in the
smooth radial distribution functions, which are due to the high number
of Fourier modes needed to fit the radial distribution function
($m=18$ in this case).  Furthermore, these oscillations do not even
agree within the 95\% confidence interval with the histogram result,
deviating especially for small distances $r$.  The explanation is that
the Jacobian factor $4\pi r^2$ in Eq.~(\ref{gij}) is still present
within the restricted sample $r<r_\mathrm{cut}$, and causes the peaks
at larger $r$ values to be fitted better than the peaks at smaller $r$
values.

\subsubsection{Overcoming over-representation through resampling}
\label{sec:resampling}

To remove the troublesome Jacobian factor in Eqs.~(\ref{gp}) and
(\ref{gij}) altogether, one can resample the inter-particle distances
found in the simulation. The idea is to draw from the original sample
of distances, $\{r_i\}$ such that smaller values of $r$ are more
likely than larger values by introducing a relative bias weight
$w(r_i)$ for each distance $r_i$ in the sample to construct a new,
resampled set $\{\tilde{r}_i\}$.\cite{footnote1b} Given
that the probability densities of the original sample points is
$p(r)=4\pi r^2g(r)/V$ [cf.~Eq.~(\ref{gp})], and provided that
subsequent resampled data points are chosen independently with weight
$w(r_i)$, the probability density of the resampled points is given by
\begin{eqnarray}
\tilde{p}(r) &=& k^{-1}w(r)p(r) =  z^{-1}r^2 w(r)g(r),
\label{tildef}
\end{eqnarray}
where $k=\int w(r)p(r)\:dr$ and $z=kV/(4\pi)$ are constants.  Note
that $z$ may be determined from
\begin{equation}
 z = \frac{V}{4\pi}\int p(r)w(r) dr = \frac{V}{4\pi}\langle{w}\rangle,
\end{equation}
which may be approximated by the average of the weight over the
samples:
\begin{equation}
  \langle{w}\rangle\approx\frac{1}{n}\sum_{i=1}^n w(r_i).
\label{simplew}
\end{equation}
In the application to the radial distribution functions below, we also
obtained $\langle w\rangle$ from a numerical integration using the
biased analytic approximation for $\tilde p$, the result of which
differed from that given by the simpler expression in
Eq.~(\ref{simplew}) by less than $0.1\%$.

\begin{figure}
  \includegraphics[width=\columnwidth]{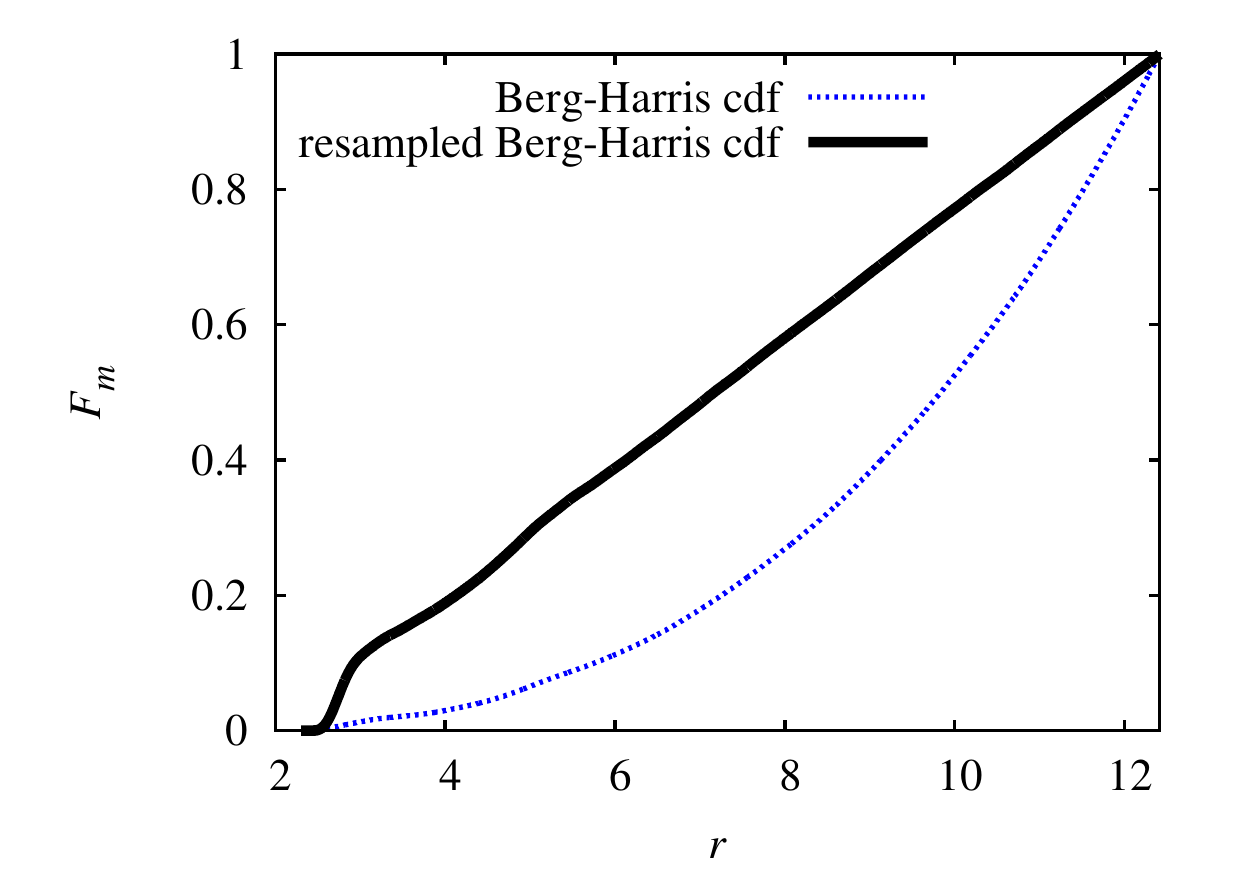} 
  \caption{The resampled and the standard 
    Berg-Harris cumulative distributions associated with the radial
    distribution function $g_\mathrm{OO}$ of the discrete water
    model.}
  \label{fig:2b}
\end{figure}

The simplest way to counter the Jacobian factor $4\pi r^2$ in
Eq.~(\ref{tildef}) is to choose a weight
\begin{equation}
w(r) = \frac{1}{r^2},
\end{equation}
which gives
\begin{equation}
g(r) = z \tilde{p}(r),
\label{show}
\end{equation}
with
\begin{equation}
z \approx \frac{4\pi}{Vn}\sum_{i=1}^n \frac{1}{r_i^2}.
\label{ceval}
\end{equation}

In detail, the resampling leading to data with a probability density
$\tilde{p}$ can be performed as follows: Given the original set
$\{r_i\}$ one determines the weight for each data point as
\begin{equation}
w_i = w(r_i) = \frac{1}{r_i^2}.
\end{equation}
One also determines the maximum weight, 
\begin{equation}
w_\mathrm{max}  = \max_{i=1\dots n} w_i
\end{equation}
to convert the weights into probabilities
\begin{equation}
p_i = \frac{w_i}{w_\mathrm{max}},
\end{equation}
which, by construction, lie between 0 and 1.  Next, one takes one of
the original sample points $r_i$ at random (with equal weight), draws
a random number $\xi$ uniformly from the interval [0,1] and if
$\xi<p_i$, one adds $r_i$ to the resampled data set
$\{\tilde{r}_i\}$. The procedure is repeated until enough resampled
points have been gathered.  There is some choice into what number of
resampled points is to be taken. In our implementation, the number of
resampled points is chosen to be the same as the number of points in
the original sample.

Note that Eq.~(\ref{show}) shows that the resampled probability
density $\tilde{p}$ and the radial distribution function $g$ are
proportional to one another. Since $g$ approaches one for large $r$,
the large values of $r$ are no longer over-represented in $\tilde{p}$.

\begin{figure}[t]
  \includegraphics[width=\columnwidth]{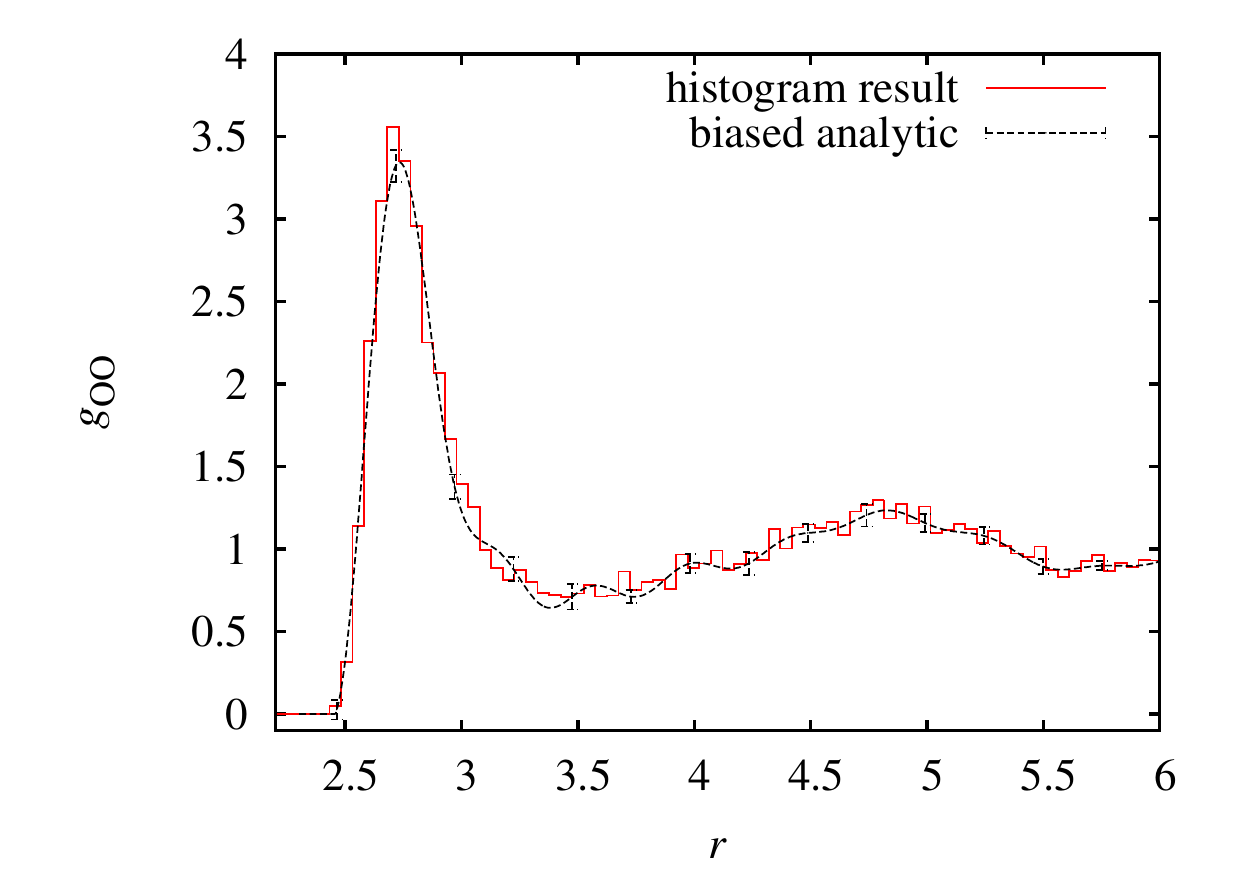} 
  \caption{The radial distribution function $g_\mathrm{OO}$ of the
    discrete water model, using the resampled smoothing method
    described.  For comparison, the results from the histogram method
    are also shown (with error bars representing 95\% confidence
    intervals at selected points).}
  \label{fig:3}
\end{figure}

For the resampled data $\{\tilde{r}_i\}$, one can use the Berg-Harris
smoothing method of Sec.~\ref{sec:spd} to obtain a smooth
approximation to $\tilde{p}$.  Fig.~\ref{fig:2b} shows the effect of
resampling on the cumulative distribution functions associated with
$g_\mathrm{OO}$ of the water model, using $Q_\mathrm{cut}=0.6$. It is
evident that the small-$r$ tail of the unweighted (Berg-Harris)
cumulative distribution is much enhanced in the resampled cumulative
distribution, and therefore more relevant to the K-S test.  Note that
we have plotted only the fits, and not the empirical cumulative
distributions, because the fits are hard to distinguish from the
empirical cumulative distributions. The differences are easier to see
in the associated radial distributions, which are plotted in
Fig.~\ref{fig:3}.  In contrast to Fig.~\ref{fig:2}, the smooth
approximation now follows the result of the histogram within error
bars for the whole range of $r$ values.  But like in Fig.~\ref{fig:2},
the supposedly smooth $g_\mathrm{OO}(r)$ exhibits fast oscillatory
behavior within those error bars.  This oscillatory behavior will be
addressed next.

\subsection{Hard-to-fit distribution functions}
\label{sec:physicalproblem}

\subsubsection{Unphysical oscillations}

It is hard to reconcile the aim of having a smooth approximation to
the radial distribution functions and the large number of Fourier
modes needed to approximate the shape of $g(r)$, which starts out as a
very flat function at small $r$, then increases sharply, reaches a
maximum not far beyond this sharp rise, and then decays on a larger
scale to $1$ in an oscillatory fashion.  The peculiar shape of $g(r)$
leads to the oscillatory behavior seen in Figures~\ref{fig:2} and
\ref{fig:3}.  Even when the oscillations lie within error bars of the
histogram results, they reintroduce a remnant of the noise similar in
magnitude to that present in the histograms that one is trying to
avoid.

It is conceivable that there are expansions other than the Fourier
expansion that are more suitable for describing a sharply increasing
function followed by a more moderate behavior.  However, it is not
clear at present what this specialized expansion should be.
Therefore, a systematic and generic way will now be presented which
avoids high-frequency modes in parts of $g(r)$ that do not require
them. This will also make the method more generally applicable to
discontinuous and other hard-to-fit densities.

We note that in the following section, the resampling explained in the
previous section is supposed to have been performed on the data
already, and that the resampled data have been sorted ($\tilde r_i\leq
\tilde r_{i+1}$). For notational convenience, the tildes on $r$, $p$
and $F$ are omitted~below.

\subsubsection{Resolving the oscillation problem using a piecewise approach}
\label{sec:piecewise}

The decomposition of $\bar F$ in Eq.~(\ref{approx}) can be adjusted to
incorporate hard-to-fit radial distributions by allowing the Fourier
decomposition to be different on sub-intervals within the total
interval $[r_1,r_n]$.  Let the full interval be divided into $k$
sub-intervals $[a_1,a_2]$, $[a_2, a_3]$, $[a_3,a_4]$, \ldots,
$[a_k,a_{k+1}]$, where $a_1=r_1$ and $a_{k+1}=r_n$.  The intervals
will be labelled by a Greek index, which can run from $1$ to $k$.  How
to choose the points $a_\mu$ (where $\mu=2\dots k$, with $a_1=r_1$
fixed) will be discussed later.  Different intervals can have
different numbers of samples that fall within its range. The fraction
of the samples that fall within a given interval $\mu$ is denoted by
$f_\mu$.

\begin{figure}[t]
  \includegraphics[width=.81\columnwidth]{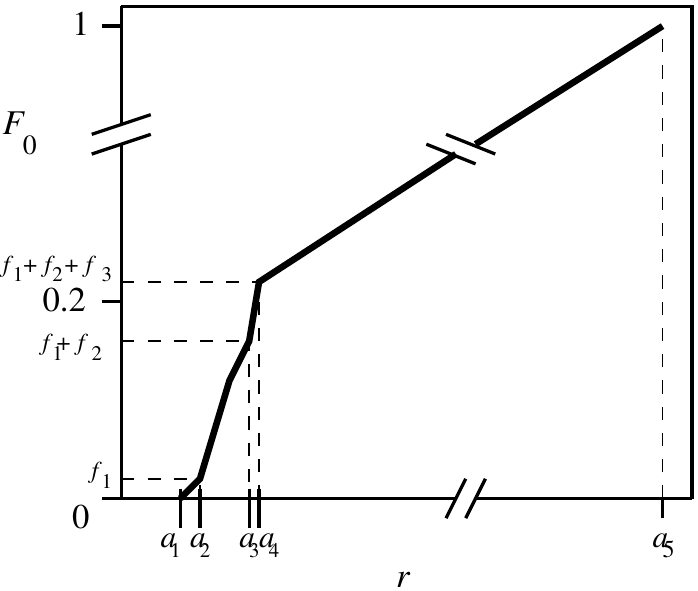}
  \caption{Illustration of a piecewise approximation $F_0$ to the
    cumulative distribution. The function $F_0$ (the solid line) is
    linear between the split points $a_1$, \dots $a_5$ (indicated by
    the dash vertical lines). The function increases by an amount
    $f_\mu$ between split points $a_ \mu$ and $a_{\mu+1}$.  The values
    of $a_\mu$ and $f_\mu$ used in this sketch roughly correspond to
    the piecewise analytic fit for the oxygen-oxygen radial
    distribution function, which requires $k=4$ intervals to achieve
    convergence.}
  \label{fig:4}
\end{figure}

\begin{figure}[t]
  \includegraphics[width=\columnwidth]{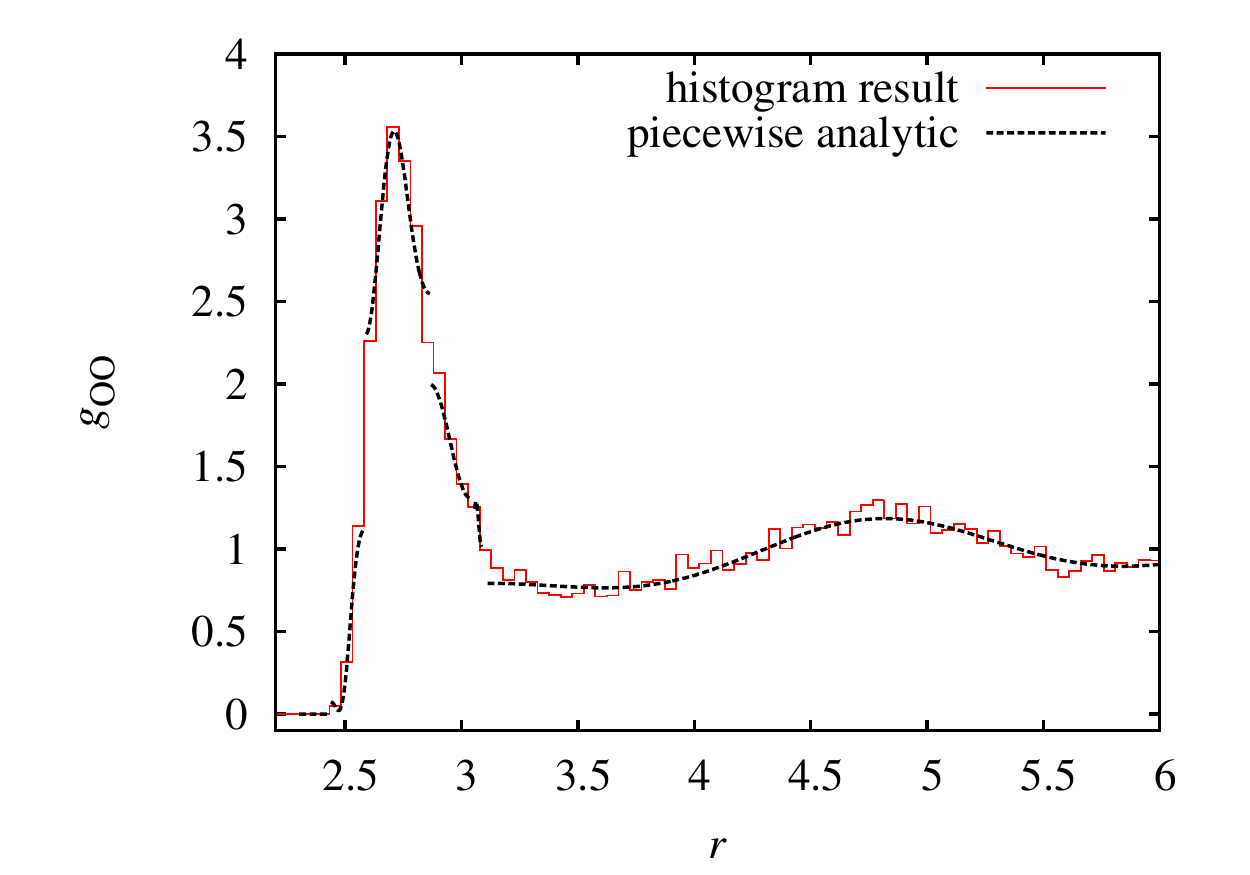}
  \caption{The radial distribution function $g_\mathrm{OO}$ of the
    discrete water model, using the first piecewise resampled
    smoothing method described in Sec.~\ref{sec:piecewise}.  For
    comparison, the results from the histogram method are also shown
    (error bars were omitted to make the comparison clearer).}
  \label{fig:5}
\end{figure}

On each sub-interval, $\bar F(r)$ is approximated by a linear part and
a truncated Fourier transform of the remainder. Analogously to
Eq.~(\ref{xdef}), the linear part in interval $\mu$ is given by
\begin{equation}
   F_{0\mu}(r) = \frac{r-a_\mu}{a_{\mu+1}-a_{\mu}} .
\end{equation}
Note that $F_{0\mu}$ ranges from 0 to 1 in the interval $\mu$.  The
piecewise linear approximation to $\bar F$ for $r$ in interval $\mu$
is then
\begin{equation}
  F_0(r) = \bar F(a_\mu) + f_\mu F_{0\mu}(r).
\label{approxinterval}
\end{equation}
where it should be noted that
\begin{equation}
\bar F(a_\mu) = \sum_{\nu=1}^{\mu-1}  f_\nu.
\end{equation}
An example of such $F_0(r)$ is shown in Fig.~\ref{fig:4}, based on
data from the oxygen-oxygen radial distribution function in the water
model (as explained below). Approximations to $\bar F$ beyond $F_0$
are found by adding Fourier modes for each interval,
\begin{equation}
  \bar F(r) \approx F_0(r) 
  + \sum_{\mu=1}^k f_\mu\chi_\mu(r)\sum_{j=1}^{m_\mu}  d_{\mu j} 
  \sin[j\pi F_{0\mu}(r)].
\label{piecewise}
\end{equation}
Here, $\chi_\mu$ is the characteristic function on interval $\mu$, i.e.,
\begin{equation}
  \chi_\mu(r) = \left\{\begin{array}{ll}\displaystyle
  1 & \mbox{ for } a_{\mu}<r<a_{\mu+1} \\
  0 & \mbox{ otherwise,}
  \end{array}\right.
\label{psijmudef}
\end{equation}
and $d_{\mu j}$ is given by
\begin{equation}
  d_{\mu j} = \frac{2}{a_{\mu+1}-a_\mu}
  \int_{a_\mu}^{a_{\mu+1}}[\bar F_\mu(r)-F_{0\mu}(r)]  \sin[j\pi
  F_{0\mu}(r)] ,
\label{ajmudef}
\end{equation}
where
\begin{equation}
  \bar F_\mu(r) = f_\mu^{-1}[\bar F(r) - \bar F(a_\mu)].
\end{equation}
The function $\bar F_\mu(r)$ is the conditional empirical cumulative
probability distribution function for data points within interval
$\mu$.  Eq.~(\ref{piecewise}) represents a piecewise analytic
approximation to the cumulative distribution.  Since the Fourier modes
form a complete orthonormal basis, the approximation becomes exact as
$m_\mu\to\infty$ for all $\mu$.  As before, however, the $m_\mu$
should not become too large to avoid spurious oscillations, which only
amounts to fitting the noise.  One therefore defines a maximum number
$m_{\mathrm max}$ of Fourier modes allowed in each sub-interval and
sub-divides intervals if the maximum number of modes is not sufficient
for convergence, as determined by the K-S test.

\begin{figure*}[t]
\centerline{\includegraphics[width=.33\textwidth]{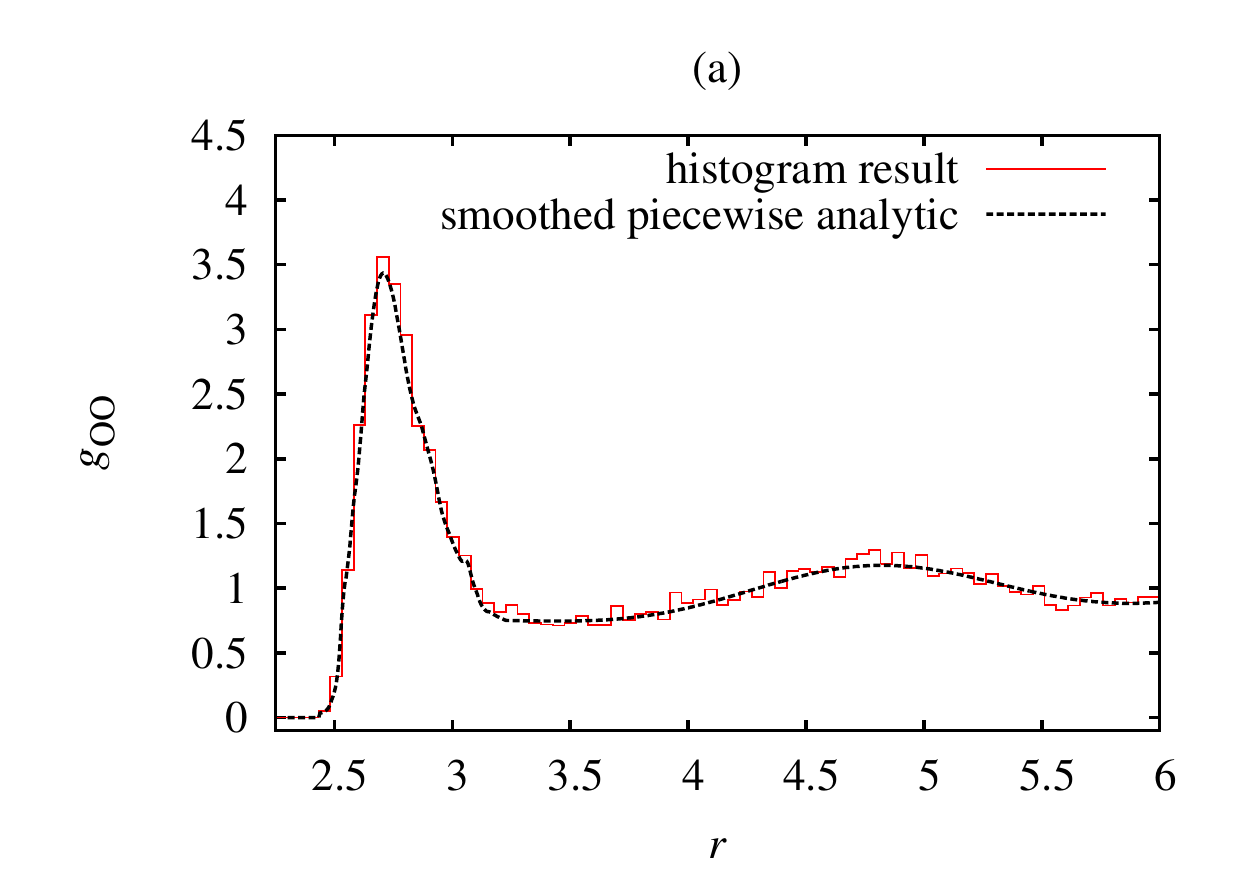}%
            \includegraphics[width=.33\textwidth]{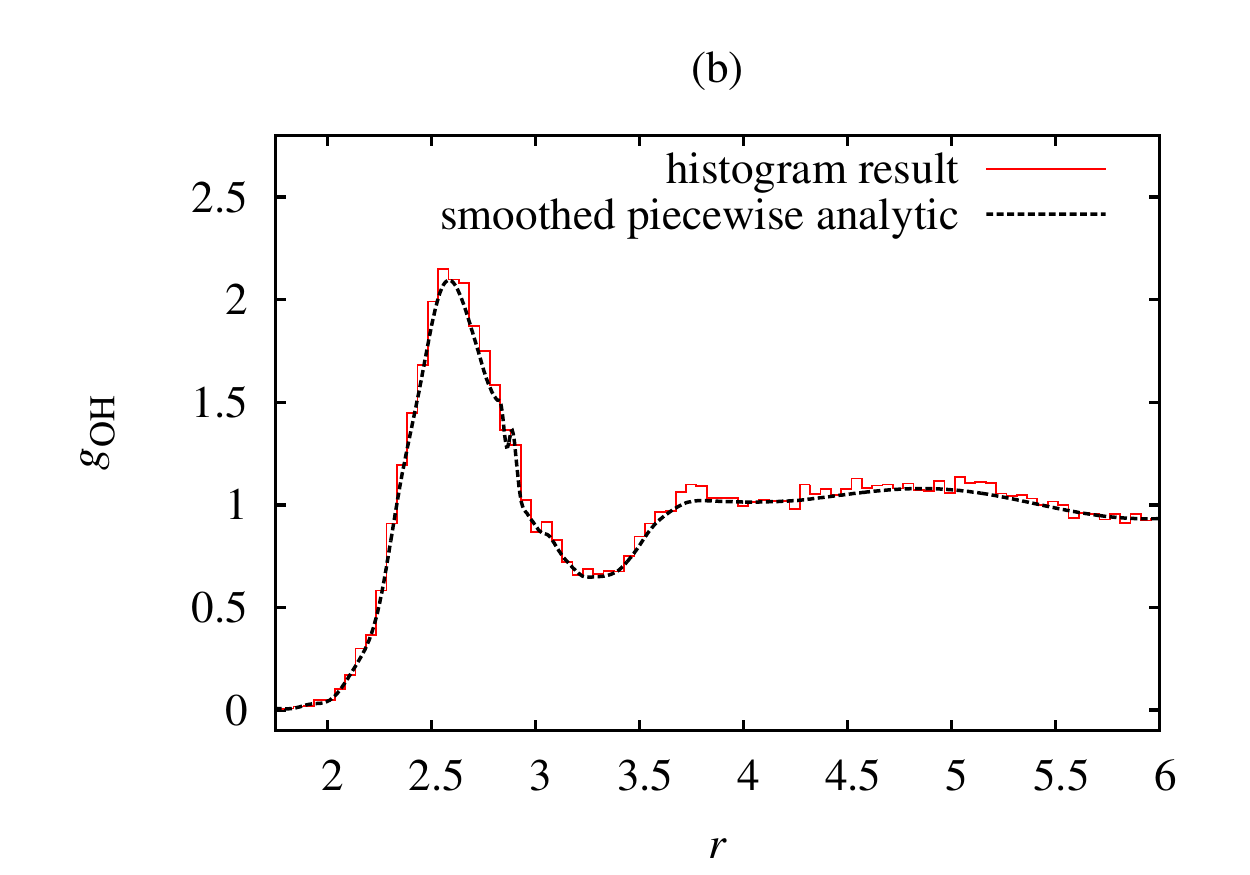}%
            \includegraphics[width=.33\textwidth]{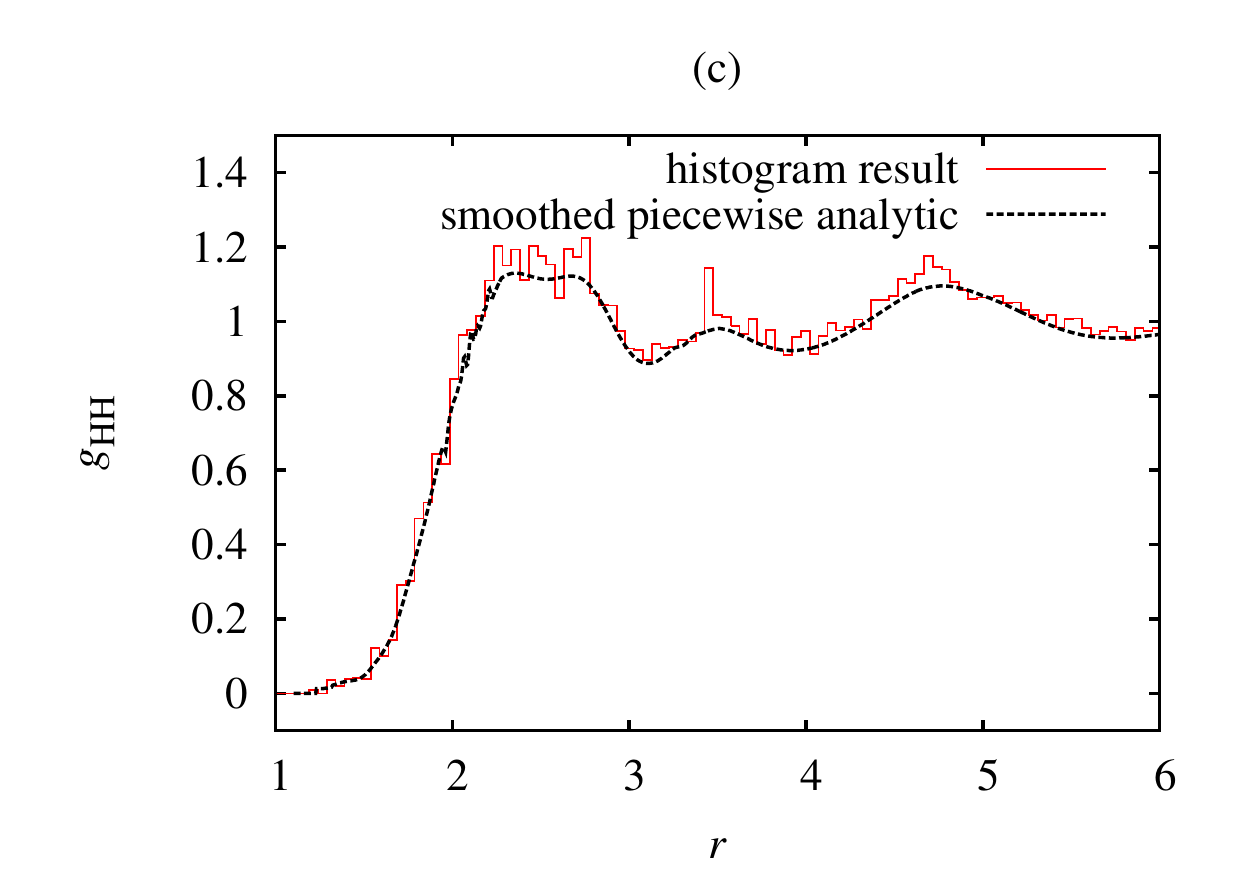}}
  \caption{The resampled, piecewise-analytic method applied to the
    radial distribution functions $g_\mathrm{OO}$ (panel a),
    $g_\mathrm{OH}$ (panel b) and $g_\mathrm{HH}$ (panel c) of the
    discretized water model, using the patched piecewise resampled
    smoothing method.  For comparison, the results from the histogram
    method are also shown, while error bars were omitted for clarity.}
  \label{fig:6}
\end{figure*}

\pagebreak[3]
The division of the original interval is chosen dynamically as follows:
\begin{enumerate}\itemsep 0pt
\item Start with one interval;
\item Increase $m$ until $Q>Q_\mathrm{cut}$;
\item If $m$ exceeds $m_\mathrm{max}$, find the most deviant point $a$
  according to the K-S test;
\item Split the interval in two at the point $a$;
\item Repeat for the conditional distribution for each interval left
  and right of $a$.
\end{enumerate}
Because the procedure is recursive, in each step the original
one-interval procedure is used, which one exception: the allowable
value of $Q$ may be set to a lower value in a sub-interval, since
there are fewer points in the interval and therefore the interval
carries less statistical weight. We have not found a unique,
statistically controlled way to adjust the $Q_\mathrm{cut}$ for
sub-intervals, but found heuristically that scaling the
$Q_\mathrm{cut}$ for interval $\mu$ by $f_\mu$ avoids fitting the
noise and leads to a satisfactory overall $Q$ value.

The recursive, multi-interval procedure is found to greatly speed up
the convergence of the approximation scheme, leading to much lower
$m_\mu$, and thus fewer oscillations.  It should be stressed that
splitting at the most deviant point $a$ in the K-S test is found to be
essential here: choosing a different splitting point does not improve
the convergence because the most deviant point is then still difficult
to fit and the $Q$ value for the sub-interval containing the most
deviant point remains the same.

Once the approximation in Eq.~(\ref{piecewise}) has been obtained, the
probability density is given by
\begin{equation}
p(r) \approx \frac{f_\mu}{a_{\mu+1}-a_\mu}\left\{1+ 
\pi \sum_{j=1}^{m_{\mu}} d_{\mu j}j \cos[j\pi F_{0\mu}(r)]\right\}
\label{ppiece}
\end{equation}
where $\mu$ is such that $r\in[a_{\mu},a_{\mu+1}]$.

The results of applying the piecewise procedure to the $g_\mathrm{OO}$
of the water model are shown in Fig.~\ref{fig:5} (remembering that
$g=zp$), with $m_\mathrm{max}=14$ and the initial $Q_\mathrm{cut}$ set
to $0.6$.  While the method now works better than without the
piecewise approach, it has one drawback: the derivative of the
approximate cumulative distribution function, which gives $g(r)$, need
not be a smooth function across the different intervals. As a
consequence, the result in Fig.~\ref{fig:5} shows artificial
discontinuities. These discontinuities at the boundaries of the
intervals fall within the 95\% confidence intervals (not shown in
Fig.~\ref{fig:5}).

\begin{figure}[b]
  \includegraphics[width=\columnwidth]{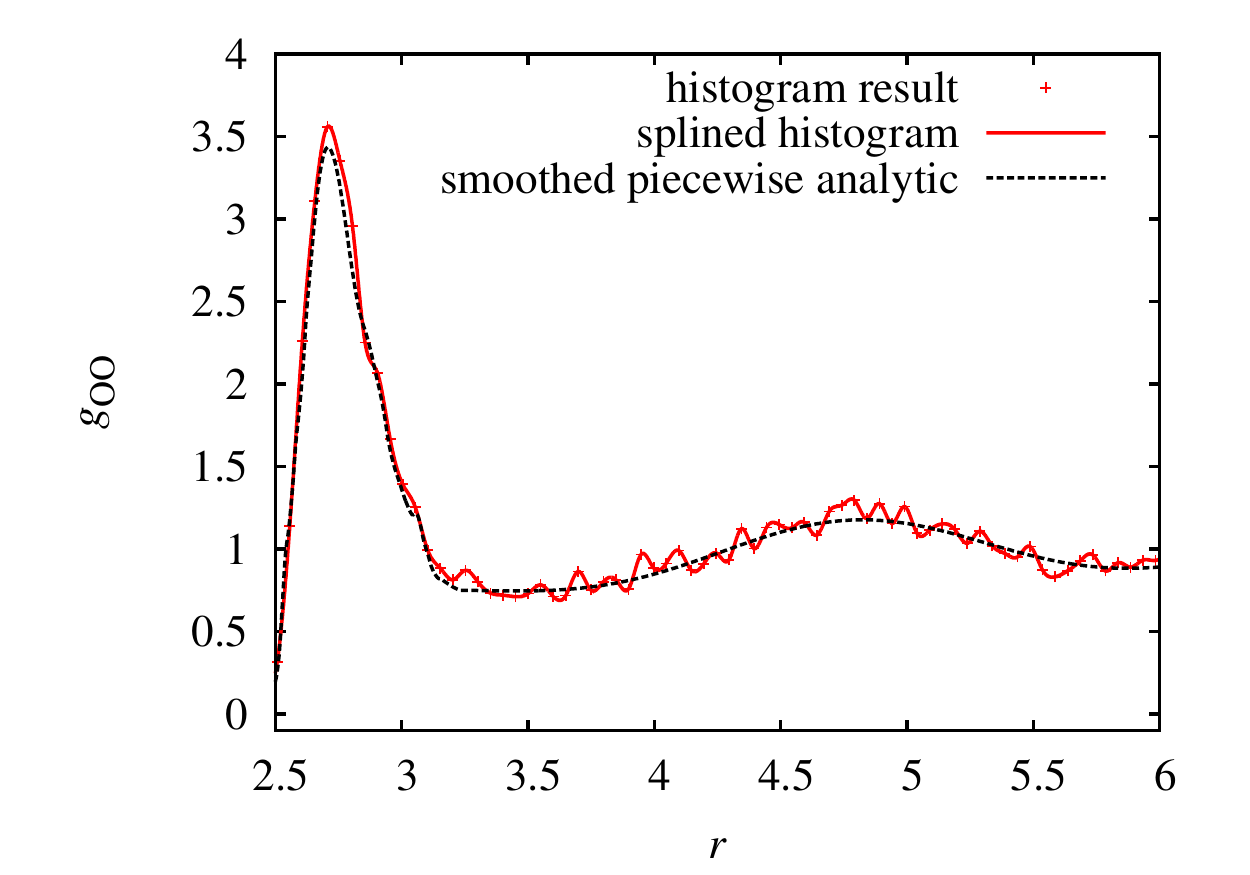}
  \caption{Comparison between the radial distribution function
    $g_\mathrm{OO}$ of the discrete water model obtained using the
    resampled piece-wise analytic method, and using a spline fit to
    the histogram results (error bars were omitted to make the
    comparison clearer).}
  \label{fig:6d}
\end{figure}

\begin{figure*}[t]
\centerline{\includegraphics[width=.33\textwidth]{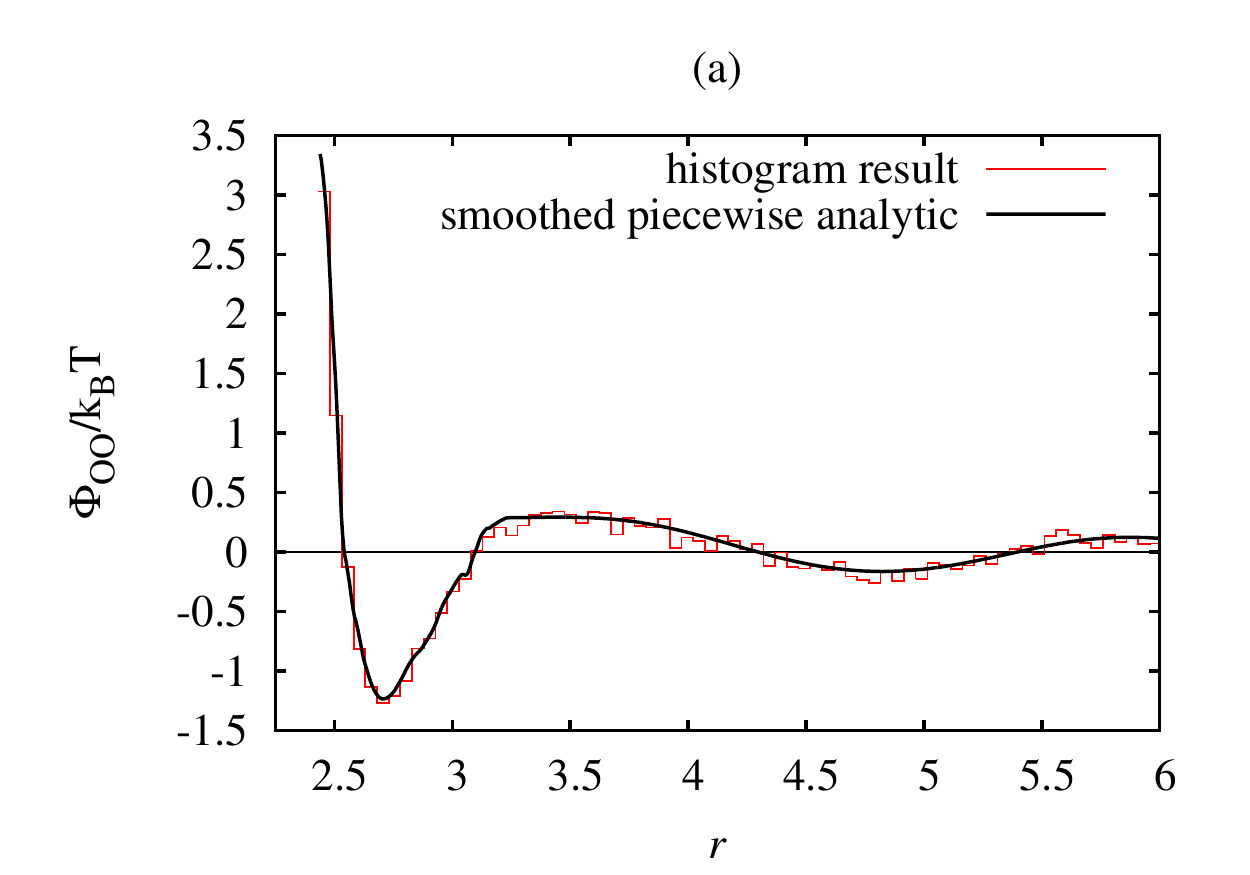}%
            \includegraphics[width=.33\textwidth]{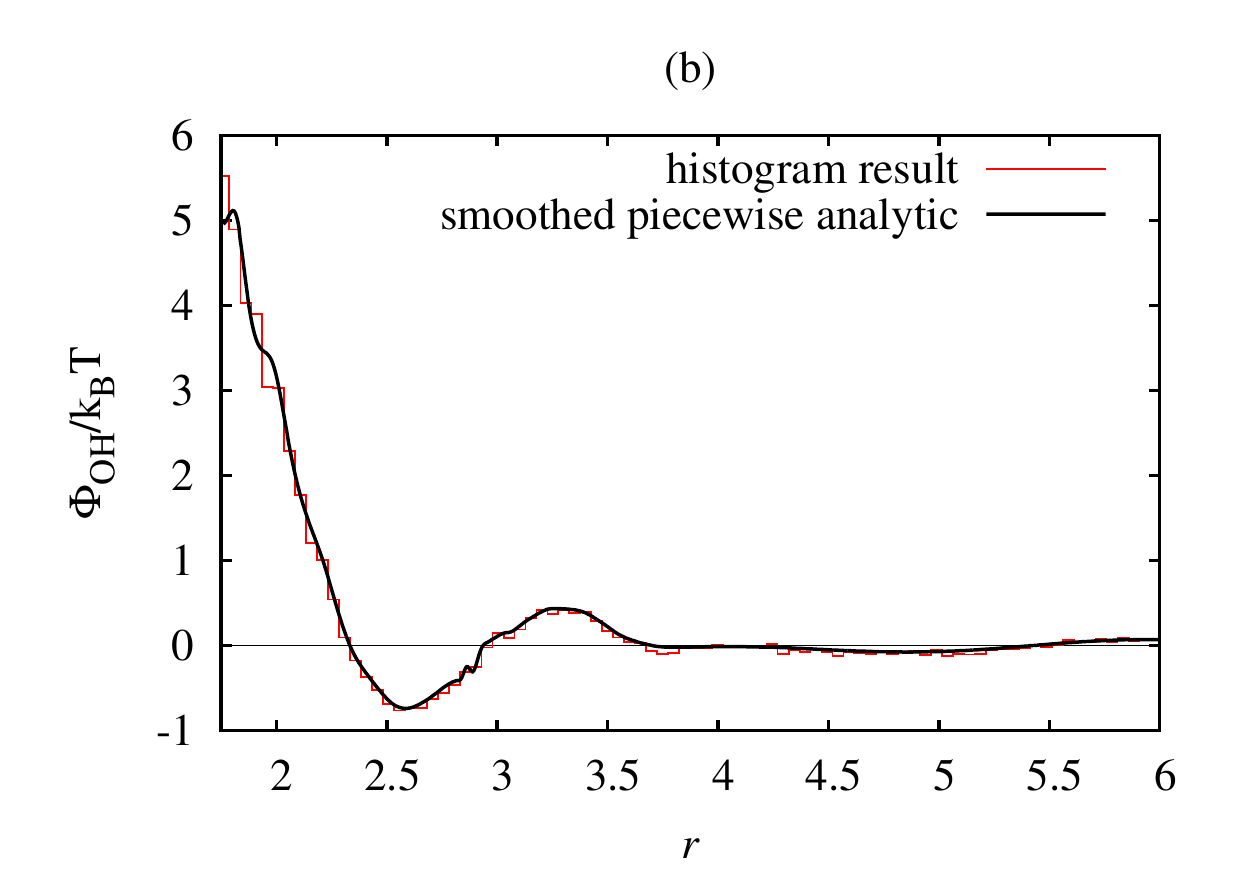}%
            \includegraphics[width=.33\textwidth]{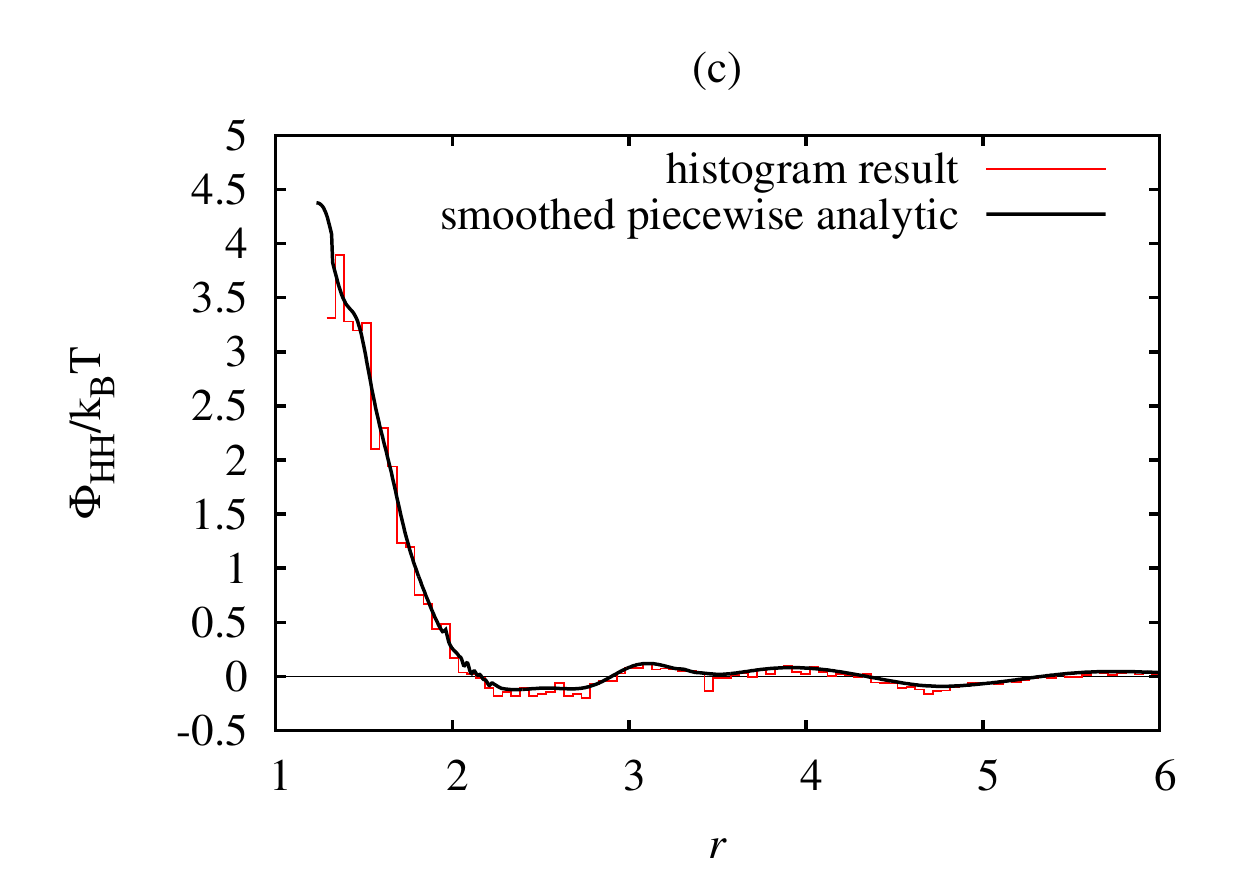}}
  \caption{The resampled, piecewise-analytic method applied to the
    potentials of mean force $\Phi_\mathrm{OO}$ (panel a),
    $\Phi_\mathrm{OH}$ (panel b) and $\Phi_\mathrm{HH}$ (panel c) of
    the discretized water model.  For comparison, the results from the
    histogram method are also shown.}
  \label{fig:7}
\end{figure*}

\subsubsection{Dealing with spurious discontinuities}

Provided the underlying probability distribution of the samples is
continuous, the spurious discontinuities that are present in the
piecewise-analytic fit will get smaller as the number of sample points
increases.  Nevertheless, for cases with poor statistics, one would
like to have an approach that gives a continuous piecewise analytic
fit to the distribution.  In fact, for some applications, having a
continuous radial distribution function is essential, such as for
Brownian simulations that take the potential of mean force as input.
Even though the discontinuities are within the statistical noise, they
would lead to sudden, unphysical changes in energy in such
applications.

There are many ways to get a continuous curve out of the discontinuous
one, but it should be remembered that one is working within the
statistical noise. We can therefore choose any method as long as the
``patch'' is still statistically reasonable. To determine the
statistical suitability, one can once again use the K-S test on the
patched fit.

We have chosen the following simple procedure: We apply a quadratic
patch function around each discontinuity at $a_\mu$,
\begin{equation}
F^\mathrm{patch}_\mu(r) = \left\{\begin{array}{ll}
b_\mu [r-(a_\mu-c_\mu)]^2 & \mbox{ if } r\in [a_\mu-c_\mu,a_\mu]\\
b_\mu [r-(a_\mu+c_\mu)]^2 & \mbox{ if } r\in [a_\mu,a_\mu+c_\mu]
\end{array}
\right. .
\end{equation}
The width $c_\mu$ of the patch is adjustable, while the prefactor
$b_\mu$ follows from the requirement that the cumulative distribution
$F+\sum_\mu F^\mathrm{patch}_\mu$ has a continuous derivative at
$r=a_\mu$. Provided the different patches do not overlap, this leads
to a value of
\begin{equation}
b_\mu = \frac{\Delta p}{4c_\mu}.
\end{equation}
where $\Delta p$ is the height of the jump in the unpatched
distribution function.

Initially, each width $c_\mu$ is set equal to half the minimum
interval size on either side of the corresponding split point $a_\mu$
to avoid overlap between the patches. The $Q$ value of the patched
distribution is then determined, and if it is not smaller than the $Q$
value for the unpatched distribution, the patch is
accepted. Otherwise, the width of the patch is reduced by a factor of
two, until the $Q$ value is acceptable.\cite{footnote2} It will be
demonstrated below that this solves the problem of spurious
oscillations and discontinuities.

\subsection{Final results}

Figs.~\ref{fig:6} show the result of the resampled, piecewise
procedure for the three radial distribution functions of the water
model, $g_\mathrm{OO}$, $g_\mathrm{OH}$ and $g_\mathrm{HH}$, with
$m_\mathrm{max}=14$ and $Q_\mathrm{cut}=0.6$.  It is clear that the
piecewise analytic fit is now smoother than the histogram, hardly
shows any oscillations and is continuous. Furthermore, while error
bars were omitted in Fig.~\ref{fig:6} for clarity, it was found that
the histogram and the patched, piecewise analytic results are in
mutual agreement within the 95\% confidence intervals.

It may be argued that plotting the histograms using bars is an unfair
way of representing the histogram results. One often takes the
histograms and applies a cubic spline fit\cite{NumRecipes} to the
results, which makes the graph seem smoother.  There is of course no
\textit{a priori} reason why the splined histogram should represent
the radial distribution function better.  For this reason, we have
shown only `unsplined' histograms so far. The comparison between the
histograms and smooth approximations in these curves should really
only be done at the mid-point of the histogram bins.  Now that the
piecewise analytic smoothing approach is fully developed, however, it
is interesting to see how it compares to a smoothed spline fit of the
histogram results. Figure \ref{fig:6d} shows a plot of these two types
of smooth results for $g_\mathrm{OO}$. The bin size for the histograms
has been chosen such that the first peak of the radial distribution is
well resolved.  One sees that the cubic spline fit does a reasonable
job for the first peak, but that in the second peak in the radial
distribution function the spline fit is still noisy, while the
statistically controlled piecewise analytic result is not.  Thus, the
splines cannot fix the roughness of the histogram method, at least not
when bin sizes are uniform.

To show how the piecewise analytic method performs for potentials of
mean force, the smooth potentials of mean force between the different
species corresponding to the radial distributions in Figs.~\ref{fig:6}
[cf. Eq.~(\ref{pmf})], have been plotted in Figs.~\ref{fig:7}.  The
results from the piecewise analytic method are considerably smoother
than the histogram results, but still exhibit some roughness since
they were based only on four configurations.

\section{Further applications}
\label{sec:furtherapp}

\begin{figure}[b]
  \includegraphics[width=\columnwidth]{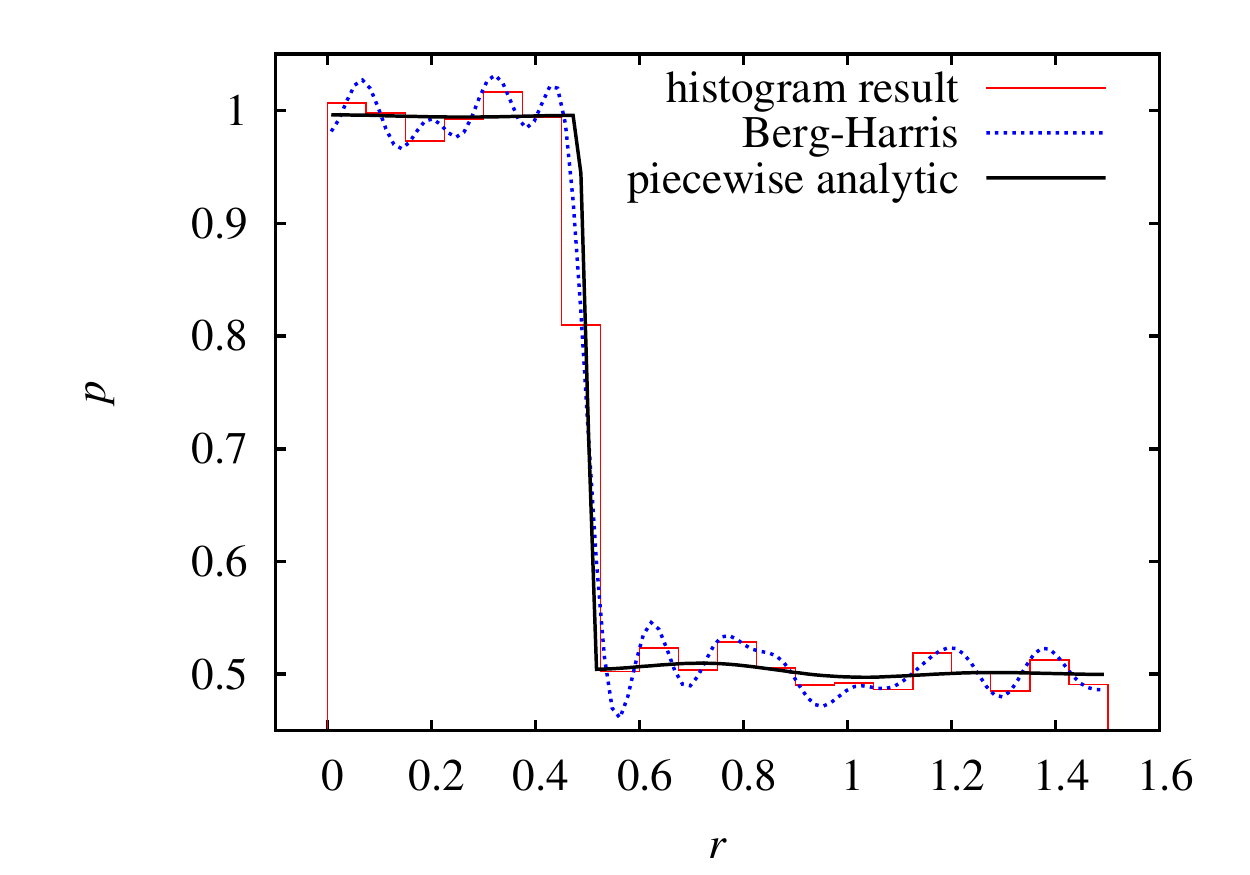}
  \caption{The piecewise-analytic method applied to $50,000$ samples
    drawn from the discontinuous distribution of Sec.~\ref{sec:disc}.
    Also shown for comparison are the results from the histogram
    method and the Berg-Harris method. For clarity, error-bars on the
    smoothing methods and the histogram have been omitted.}
  \label{fig:disc}
\end{figure}

The piecewise approximation method is not specific to radial
distribution functions, but can help to fit any probability density
that is hard to fit with a truncated Fourier series.  Below, we will
give several examples of such densities and show the advantages of
using the piecewise analytical approximation method.

\subsection{A discontinuous density}
\label{sec:disc}

Consider a random variable $r$ with a distribution $p$ given by
\begin{equation}
p(r) = \left\{\begin{array}{ll}
0 & \mbox{ if } r<0\\
1 & \mbox{ if } 0<r<\frac12\\
\frac 12 & \mbox{ if } \frac12<r<\frac32\\
0 & \mbox{ if } r>\frac32
\end{array}\right. .
\end{equation}
Note that within the domain of this function $[0,\frac32]$, there
is a discontinuity at $r=\frac12$. Discontinuities are very poorly
represented by truncated Fourier series.\cite{Jeffreys}

{}From the above distribution, 50,000 samples were drawn and used as
input to the piecewise analytic approximation (with
$m_\mathrm{max}=14$ and the initial $Q_\mathrm{cut}=0.6$), as well as
to the one-interval analytic approximation (with the same
$Q_\mathrm{cut}=0.6$ and unrestricted $m_\mathrm{max}$) and the
histogram method. The results are shown in Fig.~\ref{fig:disc}.

One clearly sees the trouble that the Berg-Harris one-interval
approximation has in capturing the discontinuity, while the histogram
is very noisy.  The piecewise analytic expansion, on the other hand,
beautifully captures the whole distribution; the automated procedure
divides the interval $[0,\frac32]$ in two at $r=\frac12$ and then
needs zero Fourier modes to approximate the separate pieces.

\subsection{The standard Cauchy distribution}
\label{sec:longtails}

\begin{figure}[t]
  \includegraphics[width=\columnwidth]{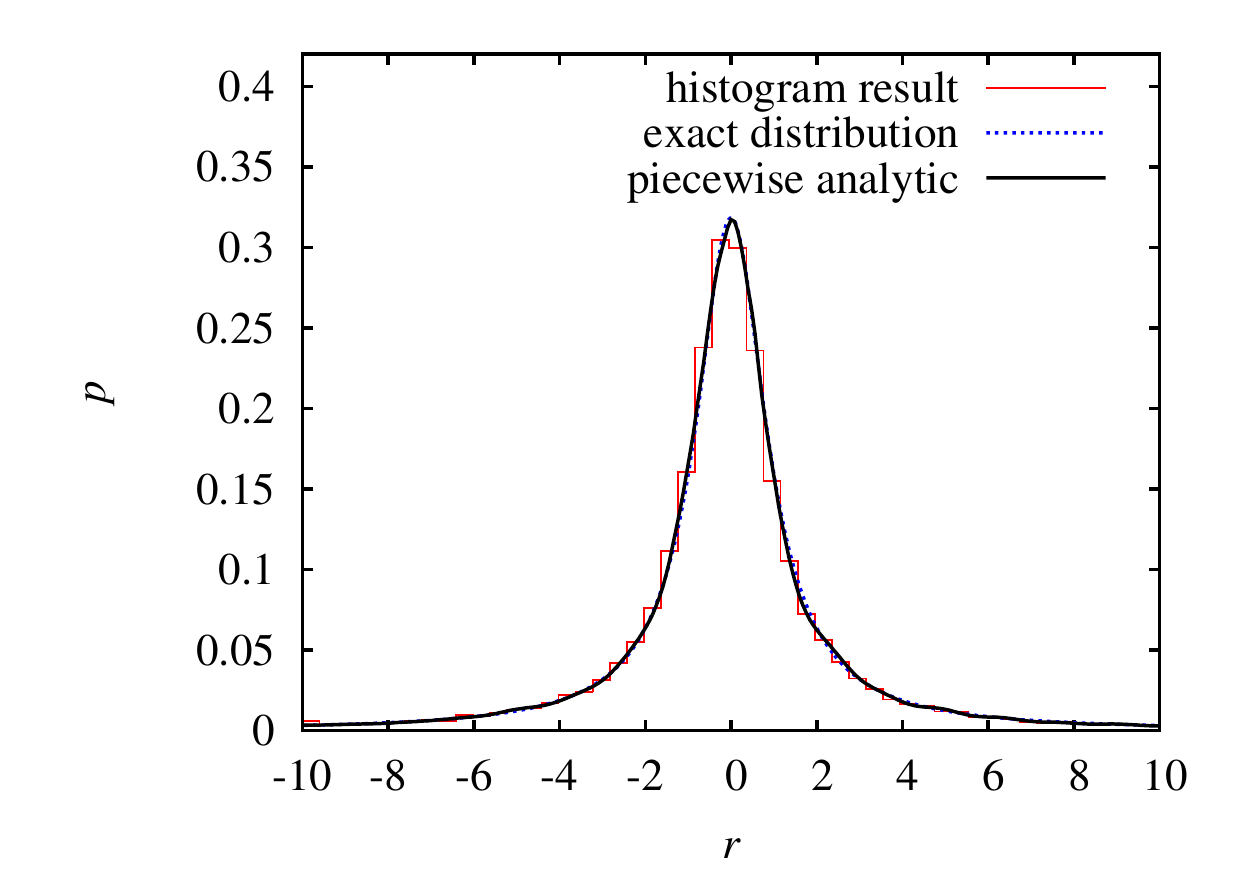}
  \caption{The piecewise-analytic method applied to $50,000$ samples
    drawn from the Cauchy distribution (cf.~Sec.~\ref{sec:longtails}).  For
    comparison, the results from the histogram method and the exact
    result are also shown. For clarity, error-bars have been
    omitted. Because the piece-wise analytic result is hard to
    distinguish from the exact distribution, the difference between
    the two is plotted in Fig.~\ref{fig:12}.}
  \label{fig:cauchy}
\end{figure}

According to Berg and Harris,\cite{BergHarris08} the original
smoothing method has trouble with densities with long tails. It is
therefore interesting to see if the piecewise approach helps for these
kinds of densities as well.  As an example, we consider the standard
Cauchy distribution
\begin{equation}
  p(r) = \frac{1}{\pi(1+r^2)}.
\label{eq:cauchy}
\end{equation}
{}From this distribution, 50,000 samples were randomly drawn and used
in the piece-wise approach (with $m_\mathrm{max}=14$ and the initial
$Q_\mathrm{cut}=0.6$). The results are contrasted with those of the
histogram in Fig.~\ref{fig:cauchy}. The piecewise smooth approximation
performs so well that it can hardly be distinguished from the Cauchy
distribution.  

Not plotted in Fig.~\ref{fig:cauchy} were the results of the original
Berg-Harris method, which, surprisingly, are so similar to the
piecewise approach that they would be hard to make out in the
figure. The only statistically significant difference between the two
results is apparent in the height of the maximum, which the
Berg-Harris fit slightly underestimates.

\begin{figure}[t]
  \includegraphics[width=\columnwidth]{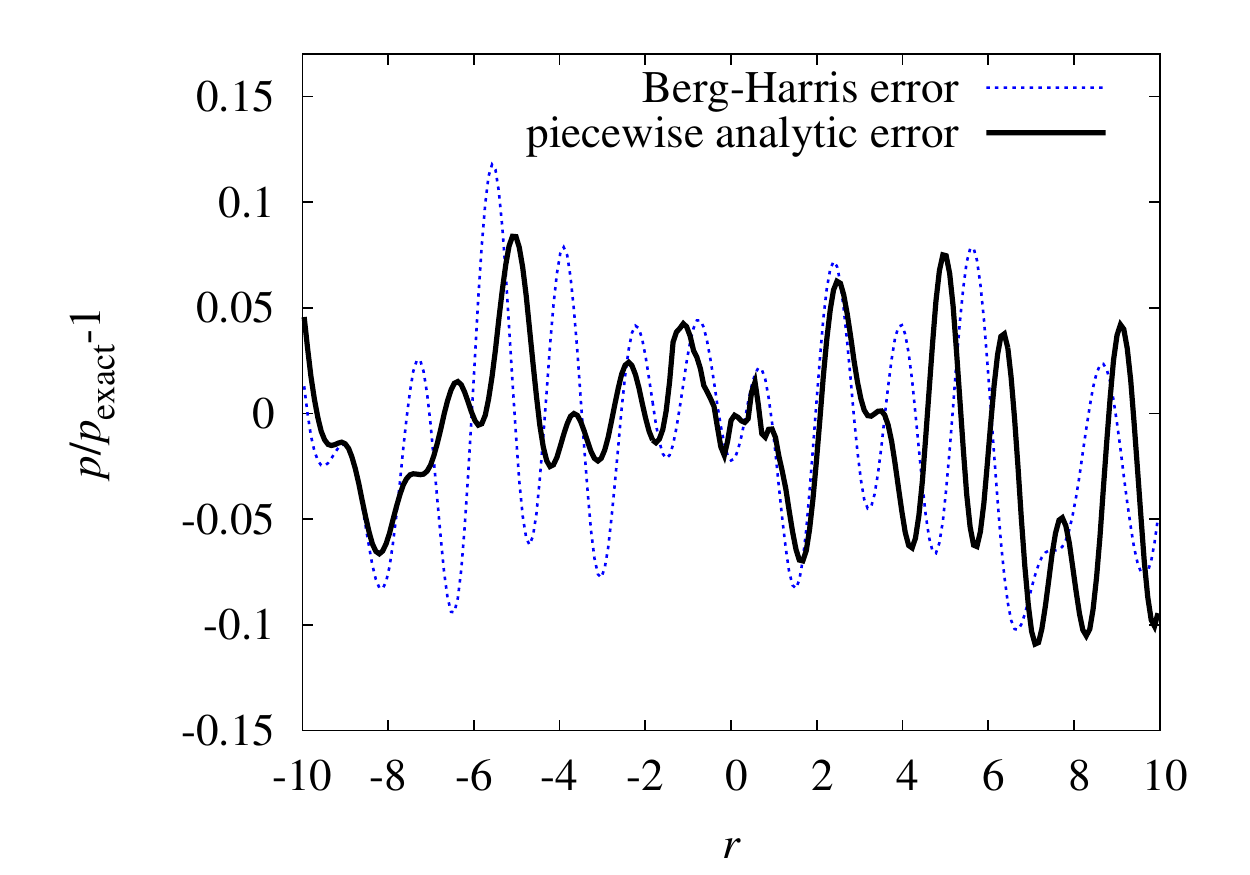}
  \caption{The error in the piecewise-analytic method applied to
    $50,000$ samples drawn from the Cauchy distribution
    (cf.~Sec.~\ref{sec:longtails}) as compared to the exact form in
    Eq.~(\ref{eq:cauchy}).  The error is here defined as the deviation
    from the exact result. For comparison, the errors in the
    Berg-Harris method are also shown.}
  \label{fig:12}
\end{figure}

The success of both
methods is illustrated further in Fig.~\ref{fig:12}, in
which the errors in the piecewise and the one-interval results are
compared; the piecewise analytic error is overall slightly smaller
than that of the Berg-Harris method, but not by much.  This success
seems to contradict Berg and Harris' warning against using the
smoothing method for densities with long tails.  It is possible the
relatively high quality of the fit in the simple approach is due to
the symmetric nature of the Cauchy density and its relative lack of
structure.

A more stringent test of the method applied to long-tailed
distributions will be presented next.

\subsection{First arrival time distribution of a diffusing particle
  to a sphere}
\label{sec:times}

Consider a particle undergoing diffusion, starting at a distance
$R_0$ from the origin, i.e. anywhere on a sphere of radius $R_0$.  The
distribution function $p(\mathbf r,t)$ of the diffusing particle satisfies
$\partial_t p=D\nabla^2p$, where $D$ the self-diffusion
constant. Investigating the time $t$ required to arrive anywhere 
on the surface of a sphere of radius $R_{min}$ for the first time is a classic case
of a first passage problem.\cite{Redner}  Such problems have
applications in the rate of molecules finding each other in a
solution, and are a major determining factor of reaction rates in
diffusion limited reactions.

It turns out that the first arrival times are distributed according
to\cite{footnote3}
\begin{equation}
  p(t) = \frac{x}{\sqrt{4\pi Dt^3}} \exp\left(-\frac{x^2}{4Dt}\right),
\label{firstpass}
\end{equation}
where $x=R_0-R_{min}$.  This distribution function has both short time
structure and a long tail.

\begin{figure}[t]
  \includegraphics[width=\columnwidth]{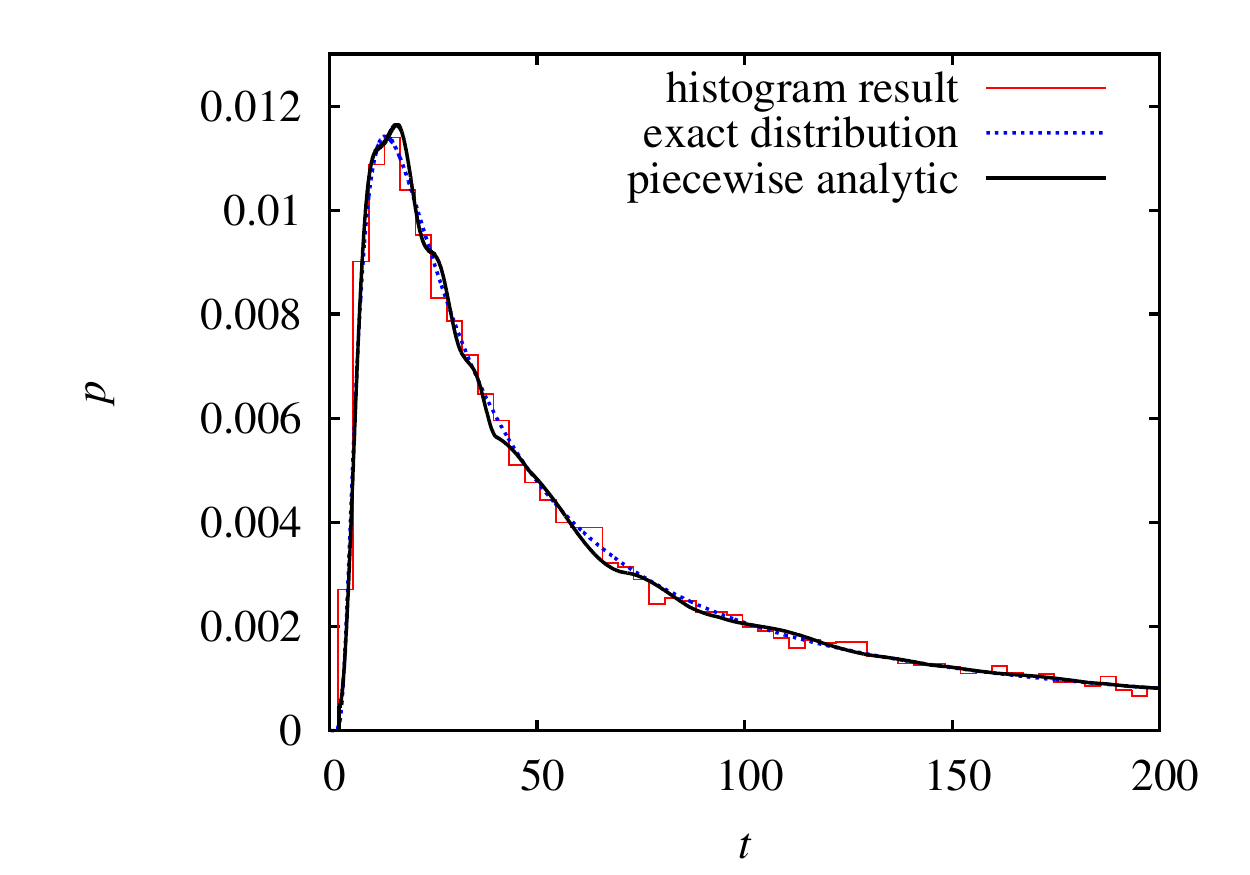}
  \caption{The piecewise-analytic method applied to $100,000$ samples
    drawn from the first passage time distribution of
    Sec.~\ref{sec:times}.  Also shown for comparison are the results
    from the histogram method and the Berg-Harris method, as well as
    the exact result.}
  \label{fig:13}
\end{figure}

{}From the distribution in Eq.~(\ref{firstpass}), 100,000 samples were
randomly drawn, with parameters set at $D=1$, $R_{min}=1$ (these
choices set the time and length units) and $R_0=10 R_{min}$. The
piece-wise approach was used to obtain an estimate of the probability
distribution (with $m_\mathrm{max}=14$ and $Q_\mathrm{cut}=0.6$), as
well as the one-interval analytic approximation ($Q_\mathrm{cut}=0.6$
and unrestricted $m_\mathrm{max}$) and the histogram method. The
results are compared in Figs.~\ref{fig:13} and \ref{fig:14}.

While all methods work to some extent, the histogram method would have
to be tailored to the function in question in order to be useful, with
differently sized bins for different values of $t$. The Berg-Harris
method works reasonably well without subdivisions, but exhibits fast
oscillations in the tails of the density, as becomes very apparent
from Fig.~\ref{fig:14}.  The appearance of oscillations is not
surprising when one considers that $m=390$ Fourier modes were needed
for the Berg-Harris fit!  The piecewise analytic result does not
exhibit fast oscillations in the tail. Furthermore, the piecewise fit
only required a total of 28 Fourier modes distributed over 8
intervals.  Although the piecewise result does have some modulation
within the error bars, we have checked that almost all of these
disappear when one takes 10 times as many data points. At that level
of statistics, the Berg-Harris results still have oscillations in the
tails, and still require over 350 Fourier modes.

When the analytical form of $p(t)$ is not known, such as for
absorption in non-spherical geometries, but samples $\{t_i\}$ are
available from numerical simulation, the piecewise approach would give
the best description of $p(t)$: one that is less noisy than the
histogram construction and with fewer oscillations than the
one-interval Berg-Harris method.

\section{Conclusions}
\label{sec:conclusions}

\begin{figure}[t]
  \includegraphics[width=\columnwidth]{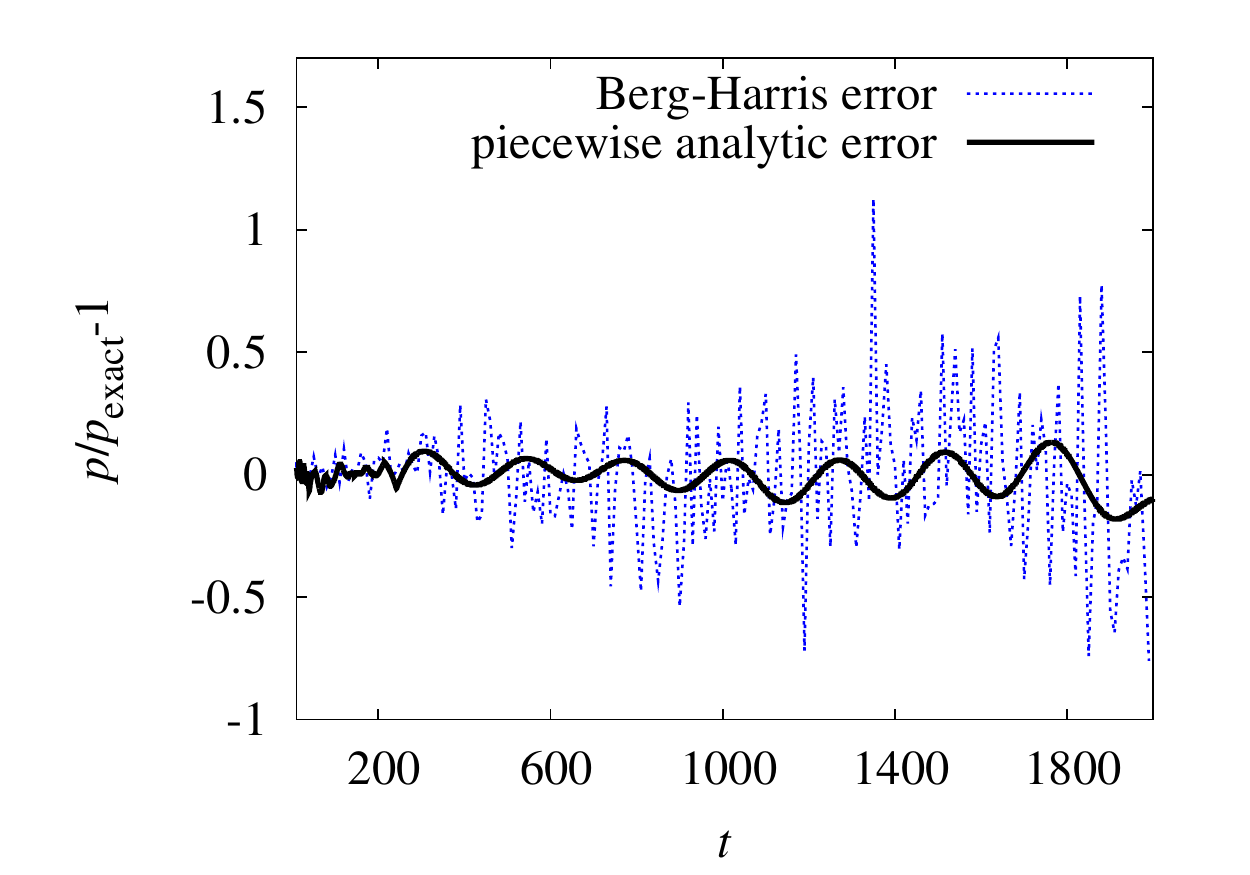}
  \caption{The relative error in the piecewise-analytic method applied
    to $100,000$ samples drawn from the first passage time
    distribution of Sec.~\ref{sec:times}, compared to the relative
    error in the Berg-Harris method.  The errors are defined here as
    the deviation from the exact result.}
  \label{fig:14}
\end{figure}

In this paper, a method to obtain smooth analytical estimates of
probability densities, radial distribution functions and potentials of
mean force in a statistically controlled fashion without using
histograms was presented. This method only uses direct samples of data
(distance samples in the case of radial distribution functions).
Since this method is expected to be generally useful, we have made our
implementation, coded in c and c++, available on the
web.\cite{footnote4}

While the method is based on the Berg-Harris method, the statistical
criterion used in that method is most sensitive to the most common
samples, which for radial distribution functions are not the ones of
physical interest.  To make the method work for radial distribution
functions, a weighted resampling of this data was required.  Spurious
oscillations, allowed by the statistical noise, were eliminated using
a piecewise approach.  In addition, one can optionally patch the
piecewise-analytic form to avoid discontinuities within the errors, if
desired.  The resampled, piecewise smoothing method was demonstrated
on data from event-driven DMD simulations of water, and proved to give
a much smoother result than the histogram method.

\begin{figure}[t]
  \includegraphics[width=\columnwidth]{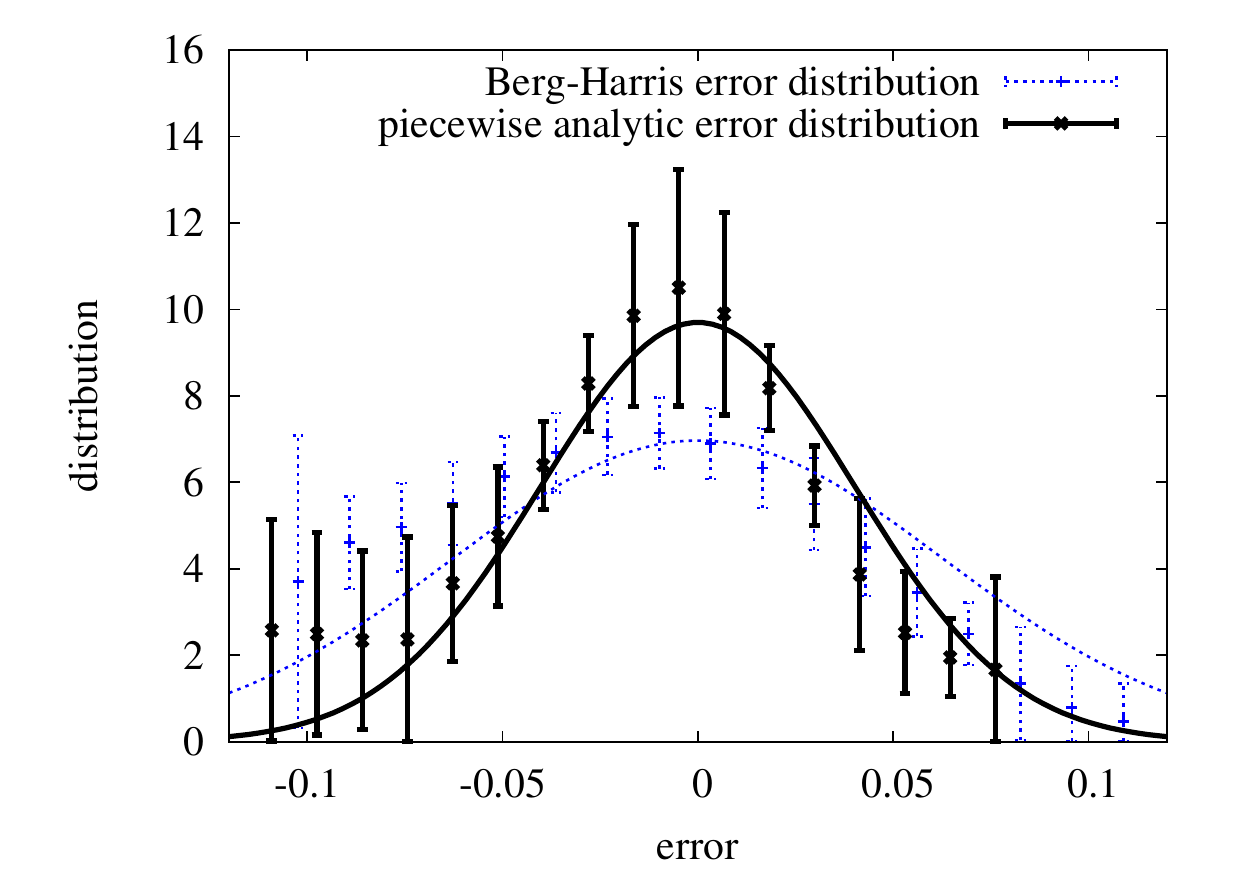}
  \caption{The distribution of the relative errors in the
    piecewise-analytic method applied to $100,000$ samples drawn from
    the first passage time distribution of Sec.~\ref{sec:times},
    compared to the distribution of the relative error in the
    Berg-Harris method.  The errors are defined here as the deviation
    from the exact result. The points with error bars are found by
    applying the piecewise analytic method to the deviations plotted
    in Fig.~\ref{fig:14} (with errors from the jackknife
    procedure), while the drawn lines are fits to
    Gaussians with a zero mean.}
  \label{fig:15}
\end{figure}

If the purpose of a simulation is just to get a good smooth
result for the radial distribution functions of a water model, one
could use the histogram method with longer simulation runs to reduce
the statistical uncertainties to some predetermined limit.
However, the histogram would still be biased and known just at a
lattice of points if a uniform bin size is used. To get a smoother and
less biased curve from the histogram, one would also have to decrease
the bin size by hand.  The piecewise analytic method makes such tuning
unnecessary. The method also make much longer runs unnecessary, since
what appears to be very poor statistics for the histogram method turns
out to be quite reasonable statistics for the piecewise analytic
method. Remember that the same set of inter-atom distance are used in
both methods.  Apparently, there is much more statistics in the sample
than the histogram is using (something that Berg and Harris also
noticed\cite{BergHarris08}).

For the test case considered here, the computational cost of doing
longer runs is not that large, so the advantages of the piecewise
analytic method may seem to be nice but not necessary.  In other
applications, however, such as in studies of the distribution of water
molecules near a polymer or a bio-molecule (or near one of their
polymer units), better statistical information is costly to obtain
because such systems are not only computationally more demanding, but
there are far less samples available in a single configuration since
only the water molecules near the polymer or the bio-molecule are
involved. The simulation run times would have to be much larger to
compensate for this poor statistics, if histograms are used. In such
cases, the piecewise analytic method is expected to be advantageous
since it appears able to use more of the information present in the
inter-particle distance sample.

The piecewise analytic method yields a smooth approximation to the
probability density function, and the deviations of the empirical
distribution from this smooth curve are supposed to be due to
statistical noise.  If this is true and the methods are unbiased, then
for large enough sample size $n$, one would expect the distribution of
errors in the probability density functions to be Gaussian with a zero
mean.  To test whether this is the case, the probability densities of
the relative errors plotted in Fig.~\ref{fig:14} were
determined, both for the Berg-Harris and for the piecewise analytic
method.\cite{footnote5} The results are shown in
Fig.~\ref{fig:15}. Within statistical uncertainty, both
distributions were found to be unbiased.  Furthermore, for the
piecewise analytic results, the error densities are roughly
Gaussian. The errors from the Berg-Harris method, on the other hand,
seem to show deviations from Gaussian behavior. The non-Gaussian
nature of the errors of the Berg-Harris method might be due to the
fact that the sample noise is fitted too closely since a large number
of Fourier modes are necessary, which means the
distribution of errors follows the distribution of the finite
sample-size errors rather than being truly random. The Gaussian nature
of the errors in the piecewise analytic method might be a confirmation
that in the smooth approximation, enough Fourier modes have be taken
into account for the remainder to be due to a random statistical
noise.

One of the nice features of the piecewise analytic method is that one
does not have to choose a bin size \textit{a priori}, as one has to do
in the histogram method. It is nonetheless true that within the
smoothing method, one is to some extent free to choose the cut-off
value $Q_\mathrm{cut}$ of the `quality' parameter $Q$ and the maximum
number $m_\mathrm{max}$ of basis functions allowed in the expansion of
each interval.  Note that both $Q_\mathrm{cut}$ and $m_\mathrm{max}$
are dimensionless numbers, and do not contain physical parameters, in
contrast to the bin size parameter required in the histogram approach.
Setting $Q_\mathrm{cut}$ too low may result in a bad fit to the data,
while setting $m_{\mathrm{max}}$ too large may result in noise fitting
and artificial oscillations.  We have found that
$Q_\mathrm{cut}\approx 0.6$ and $m_\mathrm{max}\approx 14$ are
reasonable choices, and these values were used in all the applications
presented above.

There is also some freedom in the choice of the set of basis
functions, as long as they form a complete set. The Fourier basis used
here is convenient and familiar, but others are possible.  For
instance, in Ref.~\onlinecite{VanZonetal08b}, a Chebyshev expansion
was used. This expansion was chosen over the Fourier expansion because
it gave fewer oscillations. The piecewise approach present in this
paper, however, resolves the oscillation problem as well, and is
expected to be less sensitive to the choice of basis functions than
the original Berg-Harris method.  Furthermore, the method was shown to
also be applicable to other potentials of mean force and other
probability densities with a hard-to-fit character.

\acknowledgments

We would like to thank Prof.~P.~N.~Roy for useful discussions.
This work was supported by the National Sciences and Engineering
Research Council of Canada.


\begin{thebibliography}{25}

\bibitem{McQuarrie} D.~A. McQuarrie, \emph{Statistical Mechanics}
  (Harper Collins Publishers, 1976).

\bibitem{FrenkelSmit} D.~Frenkel and B.~Smit, \emph{Understanding
  molecular simulation: From Algorithms to Applications} (Academic
  Press, San Diego, 2002).

\bibitem{Berendsen07} H.~J.~C. Berendsen, \emph{Simulating the
  Physical World: Hierarchical Modeling from Quantum Mechanics to
  Fluid Dynamics} (Cambridge University Press, Cambridge, UK, 2007).

\bibitem{BergHarris08} B.~A. Berg and R.~C. Harris,
  Comp. Phys. Comm. \textbf{179}, 443 (2008).

\bibitem{VanZonetal08b} R.~van Zon, L.~Hern\'{a}ndez de~la Pe\~{n}a,
  G.~H. Peslherbe, and J.~Schofield, Phys. Rev. E \textbf{78}, 041103
  (2008).

\bibitem{Gerickeetal08} T.~Gericke, P.~W\"{u}rtz, D.~Reitz, T.~Langen,
  and H.~Ott, Nature Physics \textbf{4}, 949 (2008).

\bibitem{Monesetal09} L.~Mones, P.~Kulh\'{a}nek, I.~Simon, A.~Laio,
  and M.~Fuxreiter, J. Phys. Chem. B \textbf{113}, 7867 (2009).

\bibitem{LiuIchiye95} Y.~Liu and T.~Ichiye, J. Phys. Chem.
  \textbf{100}, 2723 (1996).

\bibitem{ChandraIchiye99} A.~Chandra and T.~Ichiye,
  J. Chem. Phys. \textbf{111}, 2701 (1999).

\bibitem{Tanetal03} M.-L. Tan, J.~T. Fisher, A.~Chandra, B.~R. Brooks,
  and T.~Ichiye, Chem. Phys. Lett. \textbf{376}, 646 (2003).

\bibitem{FennellGezelter04} C.~J. Fennell and J.~D. Gezelter,
  J. Chem. Phys. \textbf{120}, 9175 (2004).

\bibitem{VanZonSchofield08} R.~van Zon and J.~Schofield, J. Chem.
  Phys. \textbf{128}, 154119 (2008).

\bibitem{dmd1} L.~Hern\'{a}ndez de~la Pe\~{n}a, R.~van Zon,
  J.~Schofield, and S.~B. Opps, J. Chem. Phys. \textbf{126}, 074105
  (2007).

\bibitem{dmd2} L.~Hern\'{a}ndez de~la Pe\~{n}a, R.~van Zon,
  J.~Schofield, and S.~B. Opps, J. Chem. Phys. \textbf{126}, 074106
  (2007).

\bibitem{VanZonSchofield07a} R.~van Zon and J.~Schofield,
  J. Comp. Phys. \textbf{225}, 145 (2007).

\bibitem{CyrBond07} E.~C. Cyr and S.~D. Bond, J. Comput. Phys.
  \textbf{225}, 714729 (2007).

\bibitem{AdibJarzynski05} 
 A.~B. Adib and C.~Jarzynski, 
J. Chem. Phys. \textbf{122}, 14114 (2005).

\bibitem{BasnerJarzynski08} J.~E. Basner and C.~Jarzynski,
  J. Phys. Chem. B \textbf{112}, 12722 (2008).

\bibitem{NumRecipes} W.~H. Press, S.~A. Teukolsky, W.~T. Vetterling,
  and B.~P. Flannery, \emph{Numerical Recipes in Fortran, The Art of
    Scientific Computing} (Cambridge University Press, Cambridge, UK,
  1992), 2nd ed.

\bibitem{footnote1} Alternatively, one can use other tests such as the
  Kuiper test, but these are less convenient for the piecewise
  approach presented later, since they involve several ``hard-to-fit''
  points.

\bibitem{Stephens70} M.~A. Stephens, J. Royal Stat. Soc. B
  \textbf{32}, 115 (1970).

\bibitem{Efron79} B.~Efron, The Annals of Statistics \textbf{7}, 1
  (1979).

\bibitem{Kuensch89} H.~R. K\"unsch, The Annals of Statistics
  \textbf{17}, 1217 (1989).

\bibitem{footnote1a} Another good explanation of the sensitivity of
  the K-S test to values near the median can be found in
  Ref.~\onlinecite{NumRecipes}, Ch.~4, Sec.~3.

\bibitem{footnote1b} One could imagine performing the reweighting at
  the level of K-S test instead, but this would modify the $q$
  statistic. The problem is that it would then not be known what a
  reasonable testing criterion for this new $q$ statistic should be,
  or even whether it allows a simple formulation such as $q>q_{cut}$.

\bibitem{footnote2} A procedure in which the piecewise Fourier series
  is restricted to be continuous across the interval boundaries was
  also considered but found to lead to undesirable oscillations near
  discontinuities, reminiscent of the Gibbs phenomenon.\cite{Jeffreys}

\bibitem{Jeffreys} H.~Jeffreys and B.~S. Jeffreys, \emph{Methods of
  Mathematical Physics} (Cambridge University Press, Cambridge, UK,
  1956), 3rd ed.

\bibitem{Redner} S.~Redner, \emph{A Guide to First-Passage Processes}
  (Cambridge University Press, Cambridge, UK, 1991).

\bibitem{footnote3} This is the Laplace inverse of Eq.~(6.4.8) in
  Ref.~\onlinecite{Redner}.

\bibitem{footnote4} 
  http://www.chem.utoronto.ca/\~{}rzon/Code.html.

\bibitem{footnote5} This analysis was also attempted for the
  deviations plotted in Fig.~\ref{fig:12} but the error
  bars on the resulting distributions were too large to tell whether
  they are biased or non-Gaussian.


\end{thebibliography}
\end{document}